\documentclass[ecta,nameyear,final]{econsocart}
\usepackage{zref-xr,zref-user}
\usepackage{amsmath}
\usepackage{amsthm}
\usepackage{amssymb}
\usepackage{algpseudocode}
\usepackage{caption}
\usepackage{subcaption}
\usepackage{bbm}
\usepackage{tabularx} 
\usepackage{multirow}

\RequirePackage[colorlinks,citecolor=blue,linkcolor=blue,urlcolor=blue,pagebackref]{hyperref}

\startlocaldefs

\theoremstyle{plain}

\theoremstyle{definition}
\newtheorem*{definition}{Definition}

\DeclareMathOperator*{\argmax}{arg\,max}

\endlocaldefs

\zexternaldocument*{Costly_Attention_and_Retirement_sup2}

\begin{document}

\begin{frontmatter}

\title{Costly Attention and Retirement}

\begin{aug}
%
%
%
\author[add1]{\fnms{Jamie}~\snm{Hentall-MacCuish}\ead[label=e1]{hentall-maccuish@hec.fr}}
\address[add1]{%
\orgdiv{Department of Economics and Decision Science},
\orgname{HEC Paris}}

%
\end{aug}

\begin{funding}
I thank Fabien Postel-Vinay and Eric French for discussions
and encouragement. For helpful comments, I thank Richard Blundell, Mariacristina De Nardi, Gaetano Gaballo, Tomasz Michalski, Lars Nesheim, Cormac O'Dea, Morten Ravn, Morgane Richard, Victor Rios-Rull, Arthur Seibold, Johannes Spinnewijn, and participants at the NBER SI Behavioral Macro Session, IIPF Annual Congress, RES Junior Symposium, CESifo Public Area Conference, YES, NETSPAR Pensions Workshop, Econometric Society European Meeting, CERGE-EI, ENTER, Toulouse, Cambridge, UCL, UWO, UNSW, Tilburg, Edinburgh, Royal Holloway, HEC, and the IFS.
For thoughtful discussion, I thank Andrew Caplin, Georgi Kocharkov, Dylan Moore, Leanne Nam, Nicolas Fernandez-Arias, Jim Been, Patrizia Alexandra-Massner, and Sebastian Seitz. 
Funding from Grant Inequality and the insurance
value of transfers across the life cycle (ES/P001831/1), ESRC studentship
(ES/P000592/1), and HEC is gratefully acknowledged.
\end{funding}
\begin{abstract}

In UK data, I document the prevalence of misbeliefs regarding the State Pension eligibility age (SPA) and their predictivity for retirement.
Exploiting policy variation, I estimate a lifecycle model of retirement in which, motivated by patterns in belief data, rationally inattentive households learning about uncertain pension policy endogenously generates misbeliefs.
Misbeliefs explain 51\% of the excessive (given financial incentives) drop in employment at SPA when constrained to replicate the belief data patterns and completely explain it when not. 
To achieve this, I develop a solution method for dynamic rational inattention models with persistent beliefs. 
Costly attention makes the SPA up to 15\% less effective at increasing old-age employment.
Hence, information letters improve welfare and increase employment.
\end{abstract}

\begin{keyword}
\kwd{Rational inattention}
\kwd{Retirement}
\kwd{Misbeliefs}
\kwd{Pensions} 
\kwd{Behavioral  Macro}
\kwd{Structural Econometrics}
\end{keyword}

\end{frontmatter}
\thispagestyle{empty}

\section{Introduction}

Understanding why households appear to deviate from rational behavior is crucial for policy design. 
If such deviations reflect fixed preferences or fixed features of household behavior, policy options to address them are limited. 
But mistaken beliefs about policy due to limited attention can produce similar departures, as emphasized by \cite{gabaix2019behavioral}.
In these cases, straightforward information provision might be effective.
This paper argues that misbeliefs offer an alternative, or complementary, explanation for a well-known puzzle often attributed to fixed household preferences: the disproportionately large drop in employment at pension eligibility ages, despite weak economic incentives to stop working precisely then.\footnote{This puzzle is documented in multiple countries as summarised in \cite{gruber2004social}.} 
To achieve this, I develop a general-purpose solution method for dynamic rational inattention models with persistent beliefs and use it to estimate a model on UK data, targeting both observed beliefs and behavior.

Retirement is a compelling context to study the impact of misbeliefs due to their prevalence.\footnote{Documented, for example, in \cite{gustman2005imperfect}, \cite{lusardi2011financial}, \cite{ciani2023policy}.}
Many people are confused about pensions.
In my data, 59\% of women affected by pension age reform are mistaken about their pension age by over a year when within 2-4 years of eligibility.
Initially, these misbeliefs seem strange, since the information is financially relevant and freely available.
However, they become less surprising when we acknowledge that government policy is objectively uncertain (changing in unpredictable ways), \textit{and} information is costly.
Together, policy uncertainty and costly information can generate these misbeliefs as an optimal response.
This paper asks whether these endogenously generated misbeliefs, in turn, help explain excess employment sensitivity to pension eligibility.

To investigate, I first document key facts on misbeliefs and excess employment sensitivity, then I separately and sequentially introduce policy uncertainty and information frictions (in the form of costly attention) into a model of retirement.
Specifically, I estimate a dynamic lifecycle model of retirement \citep[e.g.][]{Rust1997,French2005} with rationally inattentive households \citep[e.g.][]{Sims2003,Matejka2015,Caplin2019} deciding how much information about a changeable pension policy to acquire whilst incurring a disutility cost of information.
The model endogenously generates observed misbeliefs about the pension eligibility age, but can it generate the otherwise puzzling sharp employment drop at this eligibility age (known as State Pension Age, or SPA)?
The drop in employment at SPA is puzzling in the UK, as pension benefits are not tied to employment, so the SPA only incentivizes retirement for liquidity-constrained individuals unable to stop working earlier.
Yet, employment also falls among those with substantial liquid wealth.

Counterintuitively, unawareness of the SPA is not only consistent with high employment sensitivity to the SPA but is essential to generating this sensitivity.
The revelation of information upon reaching eligibility explains this.
In the model, households pay a utility cost to learn about their eligibility age (SPA), modeled as stochastic to capture potential government reforms.
Upon reaching the SPA, its value becomes fixed and is revealed, reflecting communication of eligibility and information disclosure during claiming.
Thus, reaching the SPA is a positive information shock.
It is also a positive wealth shock because, as households age past earlier alternative eligibility ages without receiving benefits, they rule those ages out, making now the earliest possible eligibility age.
This information shock reduces precautionary labor supply, and since leisure is a normal good, the wealth shock also reduces labor supply.
These mechanisms exist in a model with only policy uncertainty, but by introducing policy uncertainty and costly attention separately, this paper shows that the historically observed level of policy uncertainty is too low to generate meaningful changes alone.
Hence, misbeliefs generated by costly attention are key to amplifying these positive shocks at the SPA.

These model mechanisms rely on the potential for the government to reform the SPA rather than on the occurrence of reforms, though the 1995 and 2011 reforms demonstrate that this potential is real.
I do rely on the occurrence of reforms as identifying variation, firstly to estimate the probability of reform and secondly to causally identify the effect of the SPA on employment.
Support for the external validity of misbeliefs shaping claiming and retirement patterns comes from \cite{VanderKlaauw2008} and \cite{bairoliya2023revisiting}, both of which study the US, with the latter also focusing specifically on pension-policy misbeliefs.

I focus on costly attention to the SPA rather than any other burdens on people's attention for three reasons.
One, pension policy uncertainty (unlike, for example, return uncertainty) resolves, or at least diminishes, upon eligibility, making it a potential explanation for employment responses at the SPA.
Two, the SPA's simplicity (relative to other dimensions of pension-policy uncertainty, such as the benefit level) makes SPA misbeliefs easy to measure and, hence, study (additionally, this simplicity makes the misbeliefs we observe all the more surprising).
Three, it is the dimension of pension policy corresponding to the most informative subjective-belief measure in my dataset.

In the data, misbeliefs about the SPA predict employment responses to it, motivating the joint study of misbeliefs and excess sensitivity, although the direction of the correlation is initially surprising.
Women who are more mistaken about their SPA in their late 50s show a smaller response upon reaching it in their early 60s.
The model replicates this pattern because varying returns to information lead to selection into attention.
Women who do not care about the SPA neither learn nor respond upon reaching it.
So, whilst the information shock resulting from misbeliefs is an aggregate mechanism generating employment responses to the SPA, selection into SPA knowledge generates a cross-sectional correlation in which more mistaken individuals respond less.
Thus, information endogeneity and return heterogeneity are crucial for replicating the relationship between beliefs and employment.
As well as being crucial to replicating the negative correlation between misbeliefs and employment responses to the SPA, endogenous information also replicates the fact that beliefs tend toward the truth as people age.

So, endogenously chosen information  drives the relationship between retirement and misbeliefs.
Capturing this mechanism is key, but also complicates the model by introducing a high-dimensional state (prior beliefs) and choice (learning strategy).
In static rational inattention models, prior beliefs represent ex-ante heterogeneity, but in dynamic models, today's learning affects tomorrow's beliefs, making beliefs a state variable.
Many papers sidestep this by suppressing prior beliefs as a state variable.\footnote{For example \cite{miao2024dynamic,Armenter2019,turen2023state,Macaulay2021,Porcher2020}.}
While reducing the state space is beneficial and suppressing beliefs can be a good modeling assumption for specific situations, it limits the domain of application by implying that beliefs are irrelevant to choices.
It cannot capture scenarios where data show that beliefs matter and vary across individuals, such as UK pension beliefs.
Instead, I develop a solution method for dynamic rational inattention models that accommodates persistent beliefs by treating them as a state.
The method is general-purpose, as it models beliefs nonparametrically without imposing restrictions on the data-generating process.
It operationalizes the results of \cite{Steiner2017} for use with rich structural dynamic rational inattention models, while keeping the beliefs state-space representation, and addressing the computational challenges of high-dimensional state spaces by exploiting the sparsity property established in \cite{Caplin2019}.
Since beliefs are persistent, relevant, and imperfectly informed in many economically important environments, the scope for applications outside of retirement is considerable. 
For example, as the solution method makes tractable dynamic models where belief formation is an optimal response, it could be profitably applied to environments where beliefs are persistent, and their dynamics are important, such as job search (beliefs about job-finding rates) or portfolio choice (beliefs about returns).

The analysis uses data from the English Longitudinal Study of Ageing (ELSA), a micropanel survey. 
ELSA contains detailed subjective pension-belief data, in particular, self-reported and true SPAs, which allow the construction of a panel of SPA misbeliefs. 
Alongside these belief measures, it provides rich information on assets, labor-market status, and demographics. 
The survey is also linked to administrative records, in particular social-security contribution histories, enabling the estimation of individuals' State Pension entitlements. 
The recent presidential address to the Econometric Society \citep{almaas2024presidential} advocates for greater use of survey-based subjective-belief data, a call to which this paper responds.
As noted in the address, survey data allow us to observe variables, such as beliefs, that are not visible in administrative records alone.
Survey data, though, can be noisy, and I assess robustness to measurement error in beliefs data as carefully as possible. 

I estimate the model using two-stage simulated method of moments, targeting asset and employment profiles, and, when present, identifying attention costs from changes in individual misbeliefs over time.
Targeting changes in beliefs is possible thanks to my solution method, which, by retaining beliefs as a state variable, endogenously generates belief predictions that can be compared to the data.
Thus, my solution method builds a bridge between the dynamic-rational-inattention literature and the subjective-belief-data literature.
Policy uncertainty, combined with costly attention, increases the employment response to the SPA compared to a complete information baseline.
Together, they explain 51\% of the puzzling shortfall among richer households, when identifying the cost of attention from the beliefs data and completely explaining it when directly targeting the puzzle.
The mean household is willing to pay \pounds 11.00-\pounds 31.00 to learn today's SPA, so estimated attention costs are low, consistent with other evidence that apparently large deviations from optimizing behavior are explained by modest friction in dynamic environments \citep[e.g.,][]{chetty2012bounds,choukhmane_default_2025}.
Large changes in the employment response at SPA stem from small attention costs because the concentrated response at SPA partly reflects an intertemporal shift in employment relative to the frictionless benchmark.

Pension eligibility ages are considered key to increasing old-age labor force participation, which is a common policy goal.
Since costly attention increases employment response \textit{at} the SPA compared to full information, one might assume it makes the SPA a better tool for this purpose.
The opposite is generally true.
Policy experiments comparing employment increases resulting from SPA changes in versions of the model with and without information frictions show that costly attention shifts part of the informed agent's response forward but can lower the overall response.
Informed agents increase labor supply immediately, while less informed individuals, facing learning costs, respond closer to their SPA.
Thus, informing individuals, for example, by sending letters, could raise old-age employment by up to 15\%.
In most policy experiments, the benefits to households and extra tax revenue from these letters, each separately, outweigh the costs: considered jointly, information letters are always welfare-enhancing.

\paragraph*{Related Literature.\label{sec:Literature-Review}}

Dynamic lifecycle models of retirement began with \cite{Gustman1986} and \cite{Burtless1986}.
Key features introduced since then include uncertainty \citep{Rust1997}, borrowing constraints \citep{French2005}, subjective life-expectancy and Medicare \citep{VanderKlaauw2008}, and medical expenses \citep{French2011}.
Much of this literature is US-focused, and some of its concerns, like medical insurance, are irrelevant to the UK.
My model includes uncertainty, borrowing constraints, and individual heterogeneity.
\cite{ODea2018} models male UK retirees and is the closest paper in this literature.

Rational inattention began as a way to add costly attention to macroeconomic models \citep[e.g.,][]{Sims2003,Mackowiak2009, Mackowiak2015}, but now touches most fields, e.g., industrial organization \citep{brown2024endogenous}, or labor economics \citep{Bartos2016}.
\cite{Matejka2015} solve a general class of static discrete choice models with rationally inattentive agents, and \cite{Steiner2017} extends these results to dynamic discrete choice models.
A key contribution of this paper is turning the theoretical solutions of \cite{Steiner2017} into a solution method for quantitative dynamic rational inattention models with persistent beliefs.
\cite{Caplin2019} show rational inattention generically implies consideration sets, meaning solutions are sparse, which I leverage to reduce computational burden.
Dynamic rational inattention typically avoids these computational issues by suppressing the belief distribution as a state variable \citep[e.g.][]{miao2024dynamic,Armenter2019,turen2023state,Macaulay2021,Porcher2020}.
While reasonable for specific cases, this approach is not fully general and limits the range of questions that can be answered.
Two recent papers \citep{miao2022multivariate,afrouzi2021dynamic} also propose methods for dynamic rational inattention that incorporate beliefs as a state variable.
Both use the linear-Gaussian-quadratic framework popular in macro rational inattention to speed up solutions, whereas my approach handles arbitrary noise and utility but lacks the performance gains from restricting the class of models.
A closely related static rational inattention paper \cite{boehm2023intermediation} estimates a lifecycle model of older individuals, focusing on the one-shot annuity choice.

First highlighted in the US by \cite{Lumsdaine1996}, a puzzlingly large drop in employment at pension-eligibility ages is observed across countries.
In the US, the consensus was that liquidity constraints explained the drop at age 62, and Medicare eligibility the drop at age 65 \citep{Rust1997,French2005,French2011}.
Testing these explanations became possible after 2004, when the full retirement age increased.
Part of the age 65 spike followed the full retirement age, despite Medicare eligibility staying at 65 \citep{Behaghel2012}, and \cite{Mastrobuoni2009} found larger effects than standard models predicted.
Pension age increases around the world produced similar results: larger employment responses than financial incentives implied \citep[summarised in][]{gruber2004social}.
I document this in the UK, extending \cite{Cribb2016} by using richer data to rule out other potential explanations.
Part of the literature has recently converged towards reference-dependence as the explanation of this puzzle \citep[e.g.][]{Seibold2021,Lalive2017,gruber2022relabeling}.
I compare my results to this explanation in Section  \ref{sec:CompRefDep} and Online Appendix \zref{sec:RefDep}.

The use of subjective belief data in structural microeconomic models is extensive but recent (\cite{Kosar2022} provide a summary.
Most papers, however, do not model belief formation, limiting counterfactual analysis \citep[e.g.][]{de2023evaluating}.
Modeling belief formation as an optimal response to processing costs (enabled by my solution method) allows me to match model-generated beliefs to data rather than treating belief data solely as input.
Early studies of pension beliefs \citep[e.g.][]{Bernheim1988,Manski2004} document misbeliefs about benefit levels.
\cite{caplin2022communicating} find substantial misbeliefs about eligibility ages in Denmark, similar to my findings in the UK.
I use belief data to set initial conditions and identify a parameter from patterns in beliefs (patterns akin to \cite{Amin-Smith2018}, prevalent misbeliefs predicting labor supply responses, and \cite{Rohwedder2006}, errors decline as individuals age toward eligibility).

\paragraph*{Structure of the paper. }
Section \ref{subsec:Institutional-Context} provides background.
Section \ref{subsec:Data} presents the data, and Section \ref{sec:ket-facts} presents descriptive and reduced-form analysis.
Section \ref{Model} introduces the model, starting with a complete information baseline, then adding pension policy uncertainty and costly attention.
Section \ref{sec:Model-Solution} explains the solution method.
Section \ref{sec:Estimation} covers estimation.
Section \ref{sec:Results} discusses model fit and implications.
Section \ref{sec:Conclusion} concludes.

\section{Background \label{subsec:Institutional-Context}}

The UK State Pension system has changed significantly since its introduction in 1948.
I discuss the 2000-2016 system, especially post-2010, when the female SPA reform began.

\paragraph*{State Pension benefit level.}
The UK State Pension comprises two parts: the Basic State Pension, based on contributing years, and a second tier, based on earnings, both calculated over working life.
Working life is defined as spanning from the tax year an individual turns 16 to the year before they reach SPA \citep{Bozio2010}.
So, benefit entitlement is frozen a year before SPA, meaning labor supply choices near SPA do not affect the pension amount.

The Basic State Pension began in 1948.
By 2013, a full pension paid \pounds 107 per week (\$203 in 2022 USD).
Pro rata payments apply to those with fewer than 30 contributing years needed for the full pension.
Contributing years include those in the labor force (earning above a minimum threshold) and spent caring for a child or disabled person post-1978.
So, the timing of and reasons for labor market inactivity affect the pension amount.

The second tier of the State Pension began in 1978.
Initially, it used an index-linked average of earnings between lower and upper limits over working life.
Legislative changes resulted in varying accrual rates from 1978 to 2002, with a more progressive formula applied after April 2002.
Thus, the timing of earnings affects second-tier entitlements.
Private pension holders could also opt out in exchange for reduced payroll taxes.

Even in this simple outline, we see that due to protections for entitlements accrued under changing policies, the state pension benefit depends not only on total earnings and labor force participation but also on their timing and other factors \citep[see][for details]{Bozio2010}.
Still, some general trends emerge.
First, it is a relatively low benefit.
It provides a 37\% net replacement rate for median earners, compared to 47\%, 50\%, and 58\% in the USA, OECD, and EU, respectively.
Second, it is a relatively flat-rate benefit.
This is reflected in the larger drop in replacement rate between half and one-and-a-half times median earnings: 35 percentage points in the UK, versus 17, 21, and 14 in the USA, OECD, and EU \citep{OECD2011}.

\paragraph*{State Pension Age and its reform.}
The State Pension Age (SPA) is the earliest age at which the State Pension can be claimed, serving as the UK's early retirement age.
Deferring increased benefit generosity, but without a cap on deferral duration, hence implying no effective full retirement age.\footnote{Despite generous actuarial adjustments, deferral was rare, presenting a puzzle. Online Appendix \zref{sec:Extension} offers a model extension addressing this. Elsewhere, to focus on employment sensitivity, I abstract from the deferral puzzle, taking observed claiming as given.}
So, the SPA is the sole focal age of the UK state pension system.

\begin{figure}
	\flushleft
	\caption{Pension Legislation and Employment Response to the State Pension Age }
	\begin{subfigure}{0.475\textwidth}
		\includegraphics[width=0.9\linewidth]{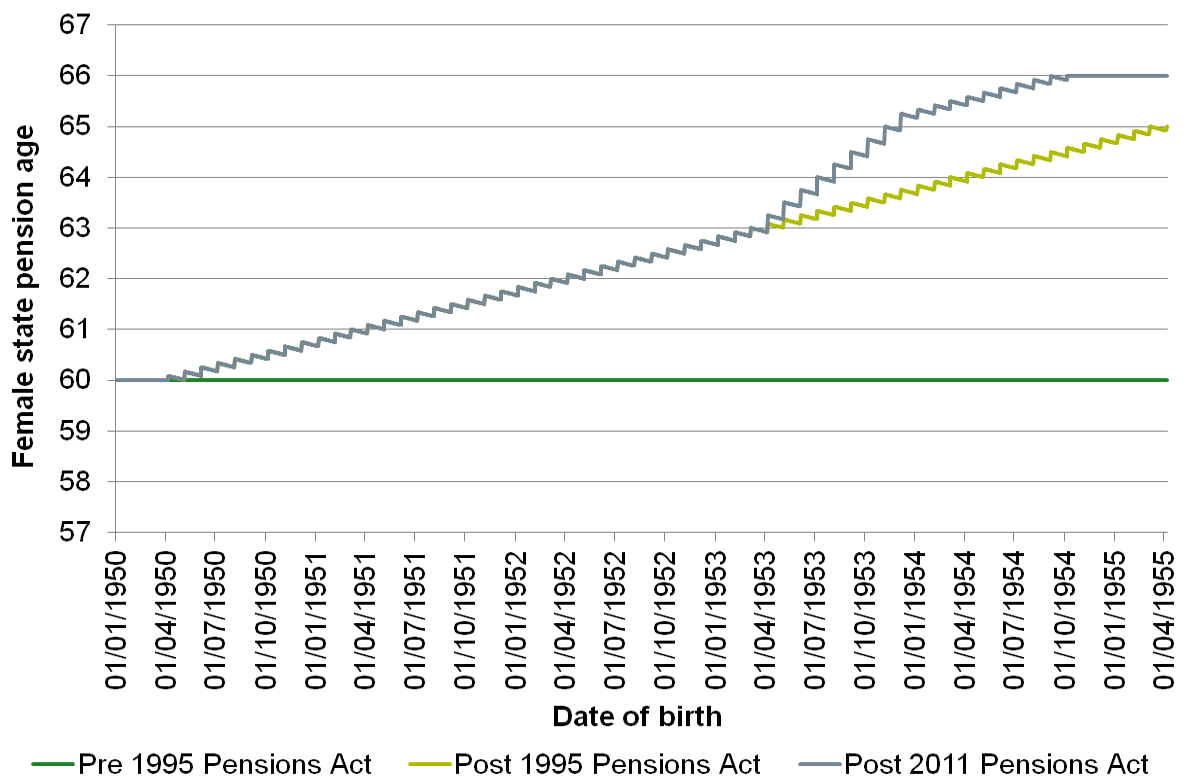}
		\caption{SPA by Date of Birth \label{fig:cribb}}
	\end{subfigure}
	\begin{subfigure}{0.475\textwidth}
		\includegraphics[width=0.9\linewidth]{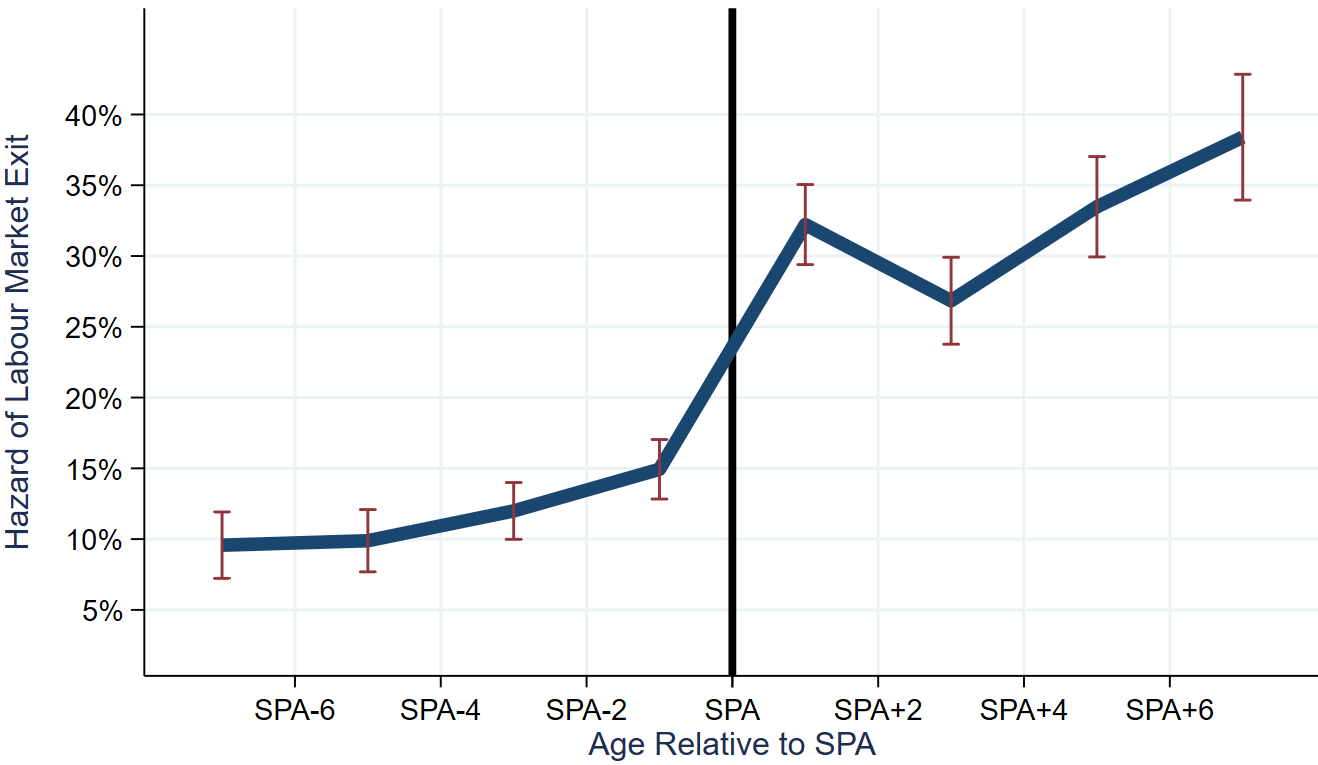}
		\caption{Fraction exiting employment\label{fig:prettyPic}}
	\end{subfigure}
	\legend{Panel (a) shows State Pension Ages for women under the Pensions Act 1995, the Pensions Act 2007, and the Pensions Act 2011. Panel (b) plots the hazard of exiting employment at ages relative to SPA, with data plotted at two-yearly intervals to match ELSA's frequency.}
\end{figure}

Unlike the State Pension amount, the SPA is a simple function of birth date and gender.
The SPA was 65 for men and 60 for women until the Pensions Act 1995, which raised the female SPA from 60 to 65 incrementally, one month every two months, over ten years starting April 2010.
The Pensions Act 2011 accelerated this change from April 2016, equalizing SPAs by November 2018, and legislated an increase for both genders to 66, phased in from December 2018.
Figure \ref{fig:cribb} shows how these changes affected women by birth cohort.
These reforms allow estimation of the risk that UK women face of SPA changes, a key model input.
I also use variation from the 1995 reform (but not the 2011 reform, to avoid confounding from a change in benefit levels) to identify the SPA's impact on employment.

\paragraph*{Communication and lack thereof.}
The government did not directly inform women affected by the reform, sending only the standard letter received by all pre-reform cohorts shortly before SPA.
This lack of communication was controversial.
From 2015, two campaign groups claimed the reforms discriminated against older women, with one unsuccessfully seeking to reverse the changes in the High Court.
Their argument focused on the lack of communication.
The government defended this by citing the absence of a national database in 1995, claiming direct notification was "essentially impossible".
Reconciling this with letter-sending at SPA is beyond this paper's scope, but the absence of protests until 20 years after legislation supports the view that reported misbeliefs are genuine.

\paragraph*{Private pensions.}
A large private pension market supplements the State Pension.
Since private pension eligibility is not tied to SPA, it has little relevance to the employment response to SPA (more evidence in Online Appendix \zref{sec:ApenEmpiricalFactors}).

\paragraph*{Excess employment sensitivity and State Pension age.}
The UK SPA reform offers a unique opportunity to examine the excess employment sensitivity puzzle, as many common explanations for labor market exits at early retirement age are ruled out.
First, UK law prohibits mandatory retirement based on age, banning it as age discrimination.\footnote{The Equality Act (2006) banned mandatory retirement below age 65, exceeding the highest SPA in this paper. The Equality Act (2010) extended the ban to all ages with exceptions in Online Appendix \zref{sec:ApenEmpiricalInst}.}
So, firm-mandated retirement cannot explain SPA employment sensitivity.
Second, the state pension is not tied to employment status; individuals can claim it and continue working, and many do.
Third, the UK pension system lacks tax incentives for labor market exits at SPA.
Unlike the US system, there is no earnings test,\footnote{Earnings tests penalize working while claiming retirement benefits, but they are \textit{not} a feature of the UK system.} and while the state pension is taxable, a component of income tax, called National Insurance contributions, is removed at SPA.\footnote{\cite{Cribb2016} find changes to participation tax rates at SPA do not explain the employment response.}
Finally, it is worth restressing that benefit entitlement is frozen the year before SPA, making it unaffected by labor supply choices near SPA.

These facts show the State Pension acts as an anticipatable increase in non-labor income, with the SPA as the eligibility age.
Announced in 1995 and starting in 2010, the reform provided at least 15 years of advance notice.
The puzzle is not that employment responds to the reform, but the concentrated response at SPA despite the long notice period.
In a standard life-cycle model with complete information and forward-looking agents, employment does not respond to anticipatable income changes unless liquidity constraints prevent intertemporal smoothing.
Liquidity-constrained individuals cannot borrow against future pension income, forcing them to wait for this income to reduce labor supply.\footnote{Loans using future pension benefits as collateral are not illegal but are not observed in practice.}
Hence, what is puzzling is the concentrated employment response among women with substantial liquid wealth.\footnote{There was also a change in means-tested benefit eligibility at SPA. Individuals are no longer eligible for working-age unemployment benefits (Jobseeker's Allowance) but become eligible for the means-tested Pension Credit Guarantee. However, with the asset cut-off threshold used to study the excess employment sensitivity puzzle, the women would not qualify for Pension Credit Guarantee, and this richer subgroup are also not major users of job-search contingent Jobseekers allowance.}

\section{Data \label{subsec:Data}}
Studying the employment response to the State Pension Age (SPA) requires a large sample of older individuals, and exploring its causes requires rich microdata.
I use the English Longitudinal Study of Ageing (ELSA), as it is the UK\footnote{ELSA \citep{ELSA} technically covers only England and Wales.} dataset best suited to these needs.

ELSA is a biennial panel dataset sampling the English population aged 50 and over, modeled on the US Health and Retirement Study (HRS).
It provides rich microdata on labor-market conditions, earnings, and asset holdings.
From wave three onward, ELSA collects data on SPA knowledge, crucial for studying misbeliefs.
ELSA requests National Insurance numbers (equivalent to a US Social Security number) and consent to link administrative records, with 80\% of respondents agreeing.
These records improve pension-entitlement estimates, which are key for modeling SPA incentives.
Survey data on health, education, and family further illuminate retirement motivations.

ELSA waves 1 (2002/03) through 7 (2014/15) cover those affected by the 1995 pension age reform and form the basis for analysis.
The main sample includes women aged 55-75 with 24,968 observations of 7,165 women.
Different samples are used only when estimating particular model inputs, such as the spousal income process (dropping females, not males) or mortality process (including older ages).
The female SPA reform began in 2010, making wave 5 the first post-reform wave.
Earlier waves control for pre-trends and inform model inputs.
The earliest affected cohort was born on 6 April 1950.
Older cohorts serve as controls and also inform model inputs.

\section{Key Motivating Facts \label{sec:ket-facts}}

\subsection{Excess Employment Sensitivity\label{subsec:First-Puzzle}}

The sensitivity of employment to official retirement ages in excess of incentive is a puzzle observed in many countries (see Section \ref{sec:Literature-Review}).
This section examines evidence of this puzzle for the UK SPA. 
The facts presented in this section are not novel, but the analysis places greater emphasis on demonstrating the puzzling nature of this employment response to SPA for standard complete information models, as this fact is central to the paper's thesis.\footnote{Indeed, \cite{Cribb2016} conduct a similar analysis. 
	They use the Labor Force Survey, which lacks sufficiently detailed asset information to rule out liquidity constraints as an explanation for the employment response to SPA. 
	Since ruling out this explanation is key to the argument in this paper, I repeat some of their analysis in ELSA, which has richer asset information.}
As liquidity constraints are the only standard complete-information mechanism for explaining SPA sensitivity (see Section \ref{subsec:Institutional-Context}), I focus on whether these constraints alone can account for the sensitivity of employment to the SPA.
Readers happy to take this puzzling nature as established will lose little by skipping to the more novel facts on the relationship between misbeliefs and the employment response to the SPA in Section \ref{subsec:Second-and-Third}.

Figure \ref{fig:prettyPic} illustrates the excess employment sensitivity puzzle, showing the mean hazard rate of exiting employment by years from SPA.
A sharp rise in exits at SPA is evident.
While this is a correlation, the female SPA reform provides policy variation with which to estimate the SPA's causal effect on labor market exit.

To do this, I use a difference-in-difference approach, common in studies of employment responses to pension eligibility \citep[e.g.][]{staubli2013does,vestad2013labour,atalay2015impact,Cribb2016,rabate2019can,etgeton2023effect}.
The outcome is the hazard of exiting employment, which captures key transitions driving employment changes and is not contaminated by the level of employment, unlike using employment directly as the outcome.
The main equation is:
\begin{equation}
	y_{it}=\alpha\mathbbm{1}\{\text{age}_{it}> SPA_{it}\}
	+ \delta_{\text{age}_{it}}
	+ \kappa_{t}
	+ \gamma_{\text{cohort}_i}
	+ \boldsymbol{\beta}^{\prime} \mathbf{X_{it}}
	+ \varepsilon_{it} \label{eq:REg},
\end{equation}
where the hazard of exiting employment ($y_{it}$) of individual $i$, at time $t$ depends on an indicator of being over the SPA ($\text{age}_{it}> SPA_{it}$); a set of quarterly age, period, and cohort dummies\footnote{As discussed in \cite{browning2012age}, these dummies flexibly absorb reduced-form variation across the three time dimensions, but separate structural age, period, and cohort trends remain set-identified.}; and a vector of controls ($X_{it}$).
The controls are marital status, years of education, highest qualification, self-reported health dummies, presence of a partner, partner's age, partner's age squared, partner's SPA eligibility, household non-housing non-business wealth (defined below), and a constant. 
I use this same set of controls throughout the regression in Section \ref{sec:ket-facts} and Online Appendix \zref{sec:ApenEmpirical}. 
The hazard ($y_{it}$) is an indicator variable defined only if the individual was employed in the previous period; it takes the value 1 if they are no longer employed and 0 otherwise.

This form assumes cohort- and date-constant age effects, age- and date-constant cohort effects, and cohort- and age-constant date effects.
Given these assumptions, which simply rephrase the parallel trends assumption, the parameter $\alpha$ is a difference-in-difference estimator of the treatment of being above the SPA.
The treatment is administered to all, but the reform induces variation in the duration of treatment.
As a pre-trends diagnostic, I test pairwise interactions among age, period, and cohort dummies and fail to reject their exclusion (age-cohort, p=0.376; age-period, p=0.144; period-cohort, p=0.666).

I prefer this simple difference-in-differences specification in the main text because it is easy to interpret, and  the final goal is to apply it as an auxiliary model for ex-post validation on the simulated data, where potential bias is not a concern. 
As long as the same biased auxiliary model is used in the data and simulations, the relevant test is replication.
For robustness, Online Appendix \zref{sec:ApenEmpiricalImp} addresses potential bias from staggered treatment timing \citep[e.g.,][]{de2020two} using the imputation method of \cite{borusyak2021revisiting}, and allowing for heterogeneous treatment effects does not materially change the conclusion.

\begin{table}
	\caption{Effects and Effect Heterogeneity of SPA Eligibility on the Hazard \label{tab:Treatment-Effect-different}}
	\centering{}
	\begin{tabular}{l*{1}{ccccccc}}
		\hline
		& (1)& (2)& (3) & (4) & (5) & (6)\\
		\hline
		{\bf Over SPA} & {\bf 0.123} & {\bf 0.088} & {\bf 0.161} & {\bf 0.150} &  {\bf 0.200 } & {\bf 0.352}\\
		\multicolumn{1}{c}{\it s.e} & (0.0235) & (0.0325) & (0.0348) & (0.0252) & (0.0459) & (0.0839)\\
		{\bf Over SPA$\times$Wealth Above Median} & \ldots & \ldots & {\bf -0.074} & \ldots & \ldots & \ldots  \\
		\multicolumn{1}{c}{\it s.e} & & & (0.0487) \\
		{\bf Over SPA$\times$Wealth in \pounds 100K } & \ldots & \ldots & \ldots & {\bf -0.026} & \ldots & \ldots  \\
		\multicolumn{1}{c}{\it s.e} & & & & (0.012) \\
		{\bf Over SPA$\times \mid$SPA - SPA Self-report$\mid$} & \ldots & \ldots & \ldots & \ldots & {\bf -0.082} & \ldots  \\
		\multicolumn{1}{c}{\it s.e} & & & && (0.0378) &  \\
		{\bf Over SPA$\times$SPA Above Self-report} &  \ldots & \ldots & \ldots & \ldots & \ldots & {\bf -0.207}  \\
		\multicolumn{1}{c}{\it s.e} & & & & & & (0.0948) \\
		\hline
		Obs. & 7,906 & 3,759 & 7,906 & 7,906  & 5,304 & 4,404\\
		\hline
	\end{tabular}
	\legend{
		The outcome variable is the hazard of exiting employment. Row 1 reports the treatment indicator for being over the State Pension Age (SPA). Rows 2–5 report interactions of this treatment indicator with: (2) an indicator for having above-median non-housing non-business wealth (NHNBW); (3) a continuous measure of NHNBW (in £100,000); (4) the absolute difference between the true and self-reported SPA; and (5) an indicator for the true SPA exceeding the self-reported SPA (i.e., under-estimation of the SPA). Self-reported SPAs are measured at age 58 or the closest available age, and wealth in the last interview before SPA.
		Each column reports estimates from a different specification of Equation \ref{eq:REg}: (1) the baseline regression on the full sample, (2) the baseline regression on an above-median NHNBW subsample, and specifications that fully interact the baseline equation with (3) being in the above-median NHNBW subsample, (4) the continuous NHNBW measure, (5) the absolute difference between the true and self-reported SPA, and (6) an indicator of under-estimating the SPA.
		Note that fully interacting the baseline specification with these variables also includes them as controls, since the interaction set contains interactions with the constant term.
		As well as the age, period, and cohort dummies, all regressions control for marital status, years of education, highest qualification, self-reported health dummies, presence of a partner, partner's age, partner's age squared, partner's SPA eligibility, household non-housing non-business wealth, and a constant.
		Following \cite{abadie2023should}, standard errors are clustered at the level of the treatment (i.e., birth cohort).
		Coefficients on controls and their interactions are not reported.
	}
\end{table}

Column (1) of Table \ref{tab:Treatment-Effect-different} presents the results of estimating Equation \ref{eq:REg}.
I find a 0.123 increase in the hazard of exiting work from being over SPA significant at the 0.1\% level.
This is in line with the findings in the literature. 
For example, studying the same reform using the Labor Force Survey \cite{Cribb2016} find an effect on employment of 6.3 percentage points with an average employment level in the range 41\%-55\%, implying an impact on the hazard rate of 11.5-15.4 percentage points.\footnote{\cite{rabate2024increasing} calculate implied impacts on the hazard rate for various papers, including this one, and arrive at slightly different numbers because they start from observed drops in employment at pension age rather than the causal estimates of the impact on retirement.}
To investigate if liquidity constraints explain the treatment effect, I restrict the sample to women from households with above-median non-housing non-business wealth (NHNBW)\footnote{I use the designation NHNBW taken from the categorization in \cite{Carroll1996}, but the variable may be better described as non-housing non-business non-pension wealth as it subtracts from ELSA's total non-pension wealth variable the value of primary residence and personal business assets. It includes current and savings accounts; shares; bonds; trusts; tax-privileged cash and equity saving instruments; other forms of land and property (excluding primary residences, farms, and personal businesses); and debt, whether credit card, personal, or other.} in the wave before reaching SPA.
The resulting threshold of \pounds 28,500 targets a group unlikely to face liquidity constraints affecting retirement choices.
As the SPA was reformed in monthly increments and Equation \ref{eq:REg} controls for quarterly age, period, and cohort effects, the control group for estimating the treatment effect consists of individuals born in the same quarter but a few months younger, thus still below SPA.
This narrow window strengthens the case against liquidity constraints: women with over \pounds 28,500 in NHNBW are unlikely to need to wait 1-3 months for the State Pension to stop working.
Column (2) of Table \ref{tab:Treatment-Effect-different} shows a treatment effect of 0.088 for this subgroup, similar to the full population and significant at 1\%.

Column (3) tests whether the treatment effects in Columns (1) and (2) differ by fully interacting Equation \ref{eq:REg} with an indicator for above-median NHNBW (i.e., allowing coefficients on all regressors, including the constant, to differ by median NHNBW group). 
The interaction with the SPA treatment is insignificant, indicating no statistically significant difference between the full-sample effect and the effect for higher-wealth households. 
As a robustness check, Column (4) replaces the binary wealth split with continuous NHNBW. 
The interaction is statistically significant but economically small: reducing the treatment effect by 1 percentage point requires an additional \pounds38,460 in NHNBW. 
Thus, while wealth matters, its effect is too small for liquidity constraints to explain the employment response to SPA.
I test the model against the two moments in Columns (1) and (2) that most directly capture the puzzle: a significant employment response to SPA, and a similarly large response among higher-wealth households.
In particular, I will focus on the model’s ability to replicate Column (2), since the response among those with liquid wealth is the puzzle.

Online Appendix \zref{sec:ApenEmpiricalExcess} shows that the wealth result is robust to more liquid asset definitions (Online Appendix \zref{sec:ApenEmpiricalAss}), and dropping controls to address bad-control concerns (Online Appendix \zref{sec:ApenEmpiricalBad}). 
Online Appendix \zref{sec:ApenEmpiricalImp} also allows for heterogeneous treatment effects using \cite{borusyak2021revisiting}. 
Across these checks, wealth affects the labor-supply response to SPA, but not enough for liquidity constraints to fully explain it.
Online Appendix \zref{sec:ApenEmpiricalFactors} considers other potential explanations for excess sensitivity, including health, private pensions, joint retirement, and employer-driven exits. 
For example, for joint retirement, Online Appendix \zref{sec:ApenEmpiricalFactors} finds no significant heterogeneity in the SPA treatment effect by whether the partner is out of work.
This is consistent with partner employment not changing discretely at one's own SPA and with evidence that spillovers from pension eligibility onto partner employment are much smaller than own effects.\footnote{See \cite{lalive2017does,johnsen2022interactions,garcia-miralles_joint_2024}.}
Similarly, these checks find little support for health or private pension eligibility as drivers since the SPA does not significantly coincide with changes in these.
As in \cite{rabate2019can}, the illegality of firm-mandated retirement helps rule out this explanation; Online Appendix \zref{sec:ApenEmpiricalFactors} further addresses the possibility of illegal mandates using self-reported reasons for employment termination.

In the rest of this paper, I use these estimates as untargeted auxiliary moments, making replication by the structural model the relevant test rather than causality. 
However, the analysis does depend on these results reflecting a puzzling response to the SPA.
Placebo tests replacing the treatment in Equation \ref{eq:REg} with indicators for aging past one or two years before SPA find insignificant effects (Online Appendix \zref{sec:ApenEmpiricalPlacebo}), confirming that the response is concentrated at SPA. This is puzzling for women with substantial liquid wealth.

\subsection{Misbeliefs and Employment Sensitivity\label{subsec:Second-and-Third}}

Compared to some other subjective belief data, such as inflation or survival expectations, pension beliefs have the distinctive feature that a currently correct answer exists. 
Hence, misbeliefs, in the sense of mistaken beliefs, are potentially observable.
This section documents such misbeliefs about the SPA that are hard to reconcile with frictionless information, since people have clear incentives to know this information.
It then investigates how these belief errors relate to observed retirement behavior.
The core finding of this section is that discrepancies in self-reported SPA predict future retirement behavior in ways consistent with costly attention.

\begin{figure}
	\flushleft
	\caption{SPA Beliefs}
	\begin{subfigure}{0.475\textwidth}
		\includegraphics[width=0.9\linewidth]{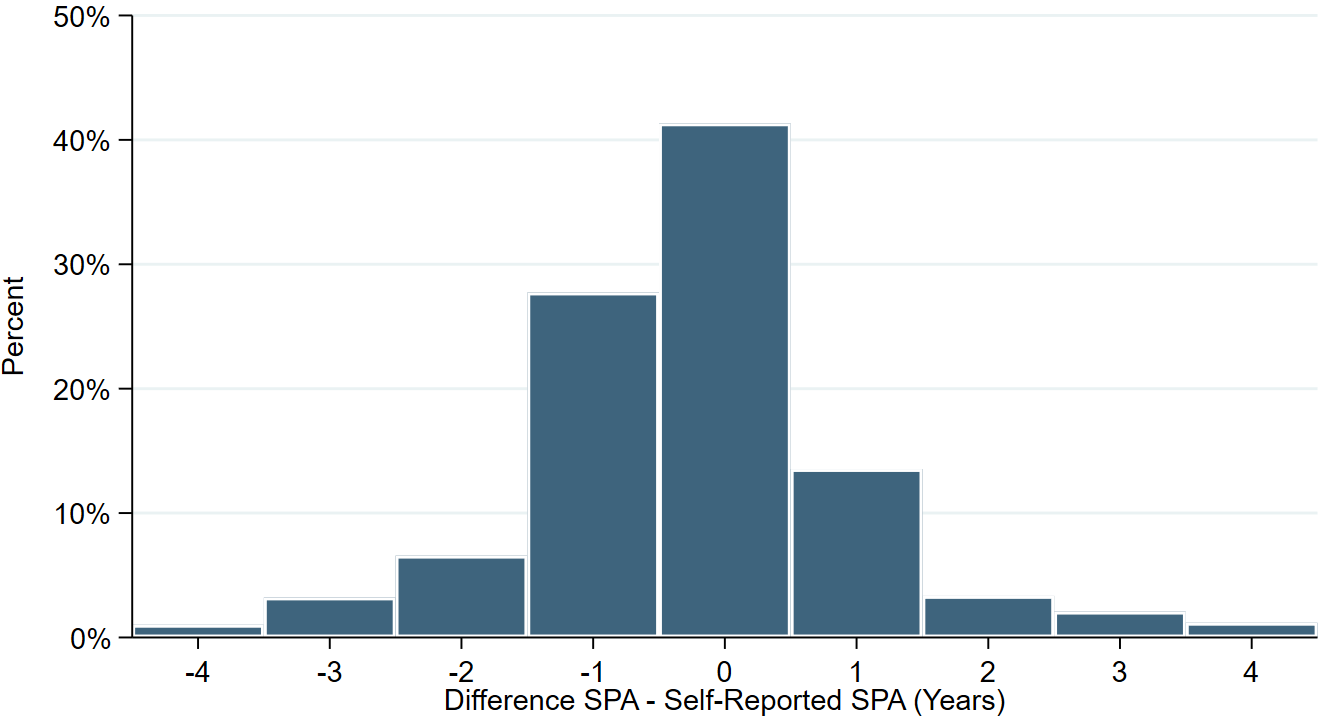}
		\caption{Mistaken SPA Beliefs Age 58\label{fig:beliefs}}
	\end{subfigure}
	\begin{subfigure}{0.475\textwidth}
		\includegraphics[width=0.9\linewidth]{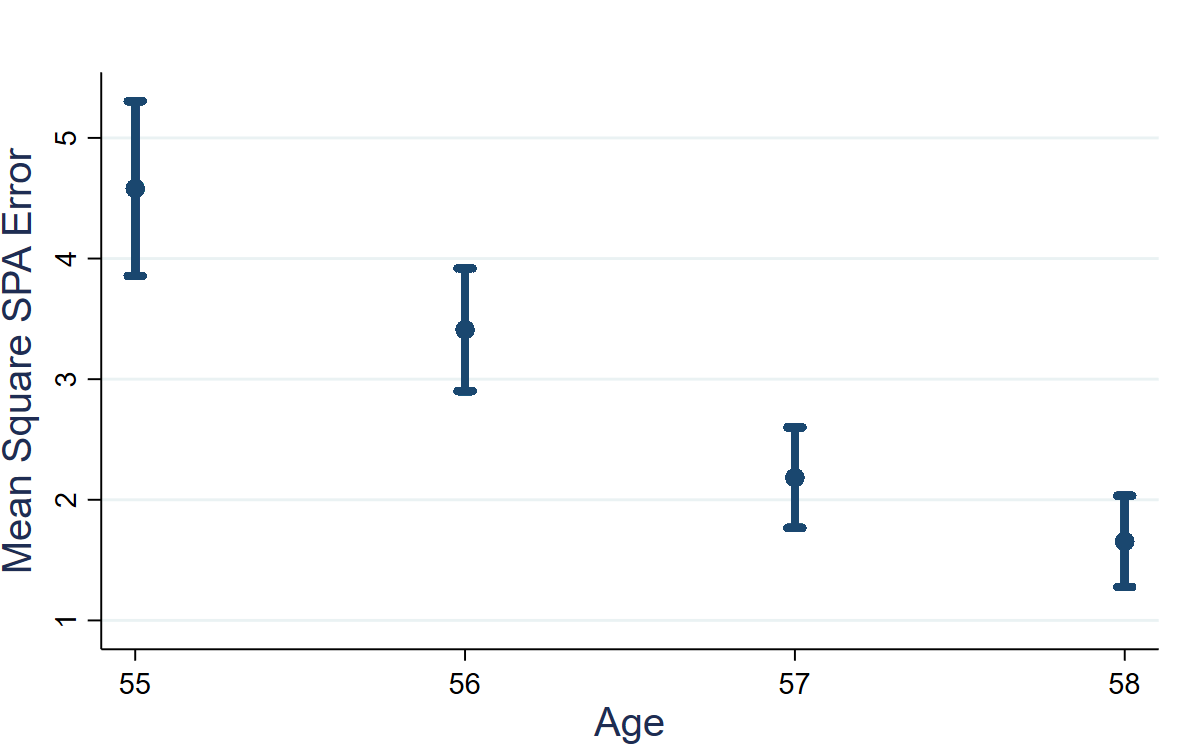}
		\caption{MSE in Self-reported SPA \label{fig:varReduc}}
	\end{subfigure}
	\legend{Panel (a) plots the frequency of errors in self-reported SPA at age 58 (binned to yearly accuracy). Panel (b) shows the mean squared error in Self-reported SPA plotted against respondents' age.}
\end{figure}

\subsubsection{Belief Data}

From wave three, ELSA asks respondents below SPA about their State Pension beliefs. 
This paper focuses on SPA beliefs in part because they provide the richest panel measure: beliefs about pension amounts are available only in wave three (so lack the panel dimension), while basic awareness of the SPA reform is too widespread to be informative (also indicating that a lack of awareness does not drive SPA belief errors). 
More details on these other belief variables are in Online Appendix \zref{sec:Other-Beliefs}.

Since ELSA has collected subjective SPA beliefs from individuals below SPA since Wave 3, we have five waves of data over almost a decade.\footnote{This reduces the sample relative to Section \ref{subsec:First-Puzzle}, since belief responses are missing in Waves 1--2 and for women already over SPA.} 
Because ELSA elicits point reports, these responses are imperfect measures of subjective beliefs \citep[e.g.,][]{de2023household}. 
Section \ref{sec:Estimation} gives them a specific model interpretation, but the analysis here requires only that they correlate with people's mean subjective SPA belief.

The question asked in the ELSA survey is: "Do you know at what age in years and months you will reach the State Pension Age?"
In addition to measurement error, this question conflates the probability of an actual reform occurring between when the person is asked and when they reach SPA with their subjective uncertainty about the current value.
If people respond precisely to this question, they would account for any expected increase in the SPA between the interview and reaching SPA.
SPA reforms, however, are observable events, and so we know the frequency with which they occur.
Hence, we can correct for the probability of a reform occurring between the interview and the respondent reaching SPA.\footnote{This assumes the individuals know the probability of reform but not the current value. It is equally possible that they know the current value but not the probability of it changing. The source of their confusion (current value or law of motion) is not separately identifiable; only the fact that they must be mistaken about one or both. Given this, I load uncertainty in this paper onto the current value, but I do not believe the implications are greatly changed if they are instead mistaken about the law of motion. }
I use a drift correction to account for the impact of possible future SPA reforms on people's responses. 
Online Appendix \zref{sec:DriftCorrect} derives the drift correction.
This drift correction does, however, impose a lot of structure, and so Online Appendix \zref{sec:ApenEmpiricalBeliefs} repeats the analysis presented below using raw responses with similar results.

Because SPA is determined by date of birth and gender, both recorded in ELSA, I can compare raw self-reported SPAs with true SPAs.
Figure \ref{fig:beliefs} plots these discrepancies for reform-affected women at age 58, or the closest age interviewed, the last age before any cohort has received an SPA communication.
Although the modal response is correct to within a year (including mistakes by a margin of months), 58.7\% are wrong by at least one year. 
Correcting for objective reform risk cannot explain these errors because the correction moves reports in the wrong direction on average, since underestimation is more common than overestimation, and is too small, at only 0.12 years on average.
Since objective reform risk cannot account for the observed discrepancies, they are best interpreted as reflecting true belief errors and measurement noise. 
Although measurement error cannot be fully isolated without an instrument, self-reported SPA discrepancies predict future retirement behavior (Section \ref{sec:EmplpoyBeliefs}) and saving behavior (Online Appendix \zref{sec:Saving-Beliefs}), suggesting they contain meaningful belief variation. 
I therefore treat drift-corrected self-reports as noisy measures of mean subjective SPA beliefs, and their discrepancy from true SPA as noisy measures of misbelief.

Misbeliefs are not only prevalent but also evolve in a way consistent with costly information acquisition.
As knowledge is retained and the value of knowing one's SPA rises with age, errors should decline with age. 
Figure \ref{fig:varReduc} supports this, showing falling mean squared errors in self-reported SPAs as women age. 
The model uses this decline to identify the attention cost.\footnote{Online Appendix \zref{sec:ApenEmpiricalDesc} shows that these discrepancies are not driven by reports of the pre-reform age or other focal ages, and also reports monthly errors and heterogeneity by education.}

\subsubsection{Relation to Employment Sensitivity \label{sec:EmplpoyBeliefs} }

This paper's model of endogenous SPA knowledge makes two distinct predictions about the relationship between SPA misbeliefs and the employment response at the SPA.
Firstly, as SPA knowledge is endogenous, selection implies that larger absolute belief errors correlate with smaller employment responses to the SPA. 
This is because if the SPA is irrelevant to an individual's actions, she will choose not to learn it or respond upon reaching it.
Secondly, because overestimating the SPA implies a positive wealth shock upon learning the actual value, it predicts that an individual who overestimates their SPA will, all else equal, have a larger employment response at the SPA than a comparable individual who underestimates.
Other mechanisms are also at play at the SPA,\footnote{Information revelation at SPA also implies a positive information shock reducing precautionary labor supply.} so the model does not necessarily predict a null or negative response among under-estimators, but it does predict a smaller one.

These two predictions are driven by distinct groups: those who care about their SPA (prediction two) and those who do not (prediction one), which are not readily observable.
Since these predictions sometimes conflict and are driven by groups whose membership is unobserved, each prediction may obscure attempts to detect the other, potentially leading to aggregation bias.
For example, if we observe an individual who massively over-predicts their SPA but does not react upon reaching it, is this evidence against the second prediction, or is this confirmation of the first?

The data robustly support the first prediction.
Column (5) of Table \ref{tab:Treatment-Effect-different} shows the results of Equation \ref{eq:REg} fully interacted with the absolute error in self-reported SPA at age 58 or the nearest age observed (i.e., interacting all regressors, including the constant in Equation \ref{eq:REg}, with the absolute error).
The significant negative interaction suggests that for each additional year of error in SPA self-reporting, the employment response drops by 8.2 percentage points.
So, those least informed about the SPA before age 60 have the smallest employment response upon reaching SPA after 60.
This pattern aligns with a model of endogenous costly information acquisition: individuals who place lower value on knowing their SPA acquire less information and exhibit smaller behavioral responses. In contrast, such a selection would not arise if self-reports were purely measurement error or if information arrived exogenously. 
Online Appendix \zref{sec:ApenEmpiricalBeliefs} addresses measurement error using squared misbeliefs, for which the bias can be signed, and a reversed regression based on the observed SPA employment response. These checks support selection into SPA knowledge as the source of the correlation between earlier misbelief size and later employment responses.

For the second prediction, the data provide stronger support after addressing aggregation bias from individuals who select out of SPA knowledge. 
To do that, I restrict the sample to those with absolute errors of at most 5 quarters, since larger errors are more likely to reflect such selection.
Column (6) of Table \ref{tab:Treatment-Effect-different} then tests heterogeneity by whether individuals under-estimate their SPA at 58 or the closest age observed by fully interacting all regressors (including the constant term) in Equation \ref{eq:REg} with an indicator for under-estimating. 
Consistent with prediction two, those who overestimate their SPA respond significantly more at SPA. 
Without the sample restriction, the point estimate has the same sign but is not statistically significant (Online Appendix \zref{sec:ApenEmpiricalBeliefs}), although measurement error in this threshold variable regression will bias the estimate toward zero \citep{aigner1973regression}.
Online Appendix \zref{sec:ApenEmpiricalBeliefs} shows that the conclusions of this section are similar without the drift correction, when including treatment heterogeneity by the size and direction of belief errors jointly, and dropping controls. 
These checks strongly support prediction one and provide noisier, but directionally consistent, support for prediction two.

Recent work explains excess employment sensitivity using reference-dependent preferences \citep[e.g.,][]{Seibold2021,Lalive2017}. This paper emphasizes a different channel: attention frictions can leave individuals uncertain about their eligibility age. 
The misbeliefs documented here, and their predictive relationship with employment responses, are difficult to explain with fixed preference parameters alone and are consistent with information frictions playing an independent role.
Section \ref{sec:CompRefDep} provides further comparison between these two explanations.

Regarding the external validity of the mechanism, it is worth emphasizing that although I use the occurrence of the reform to identify variation, the proposed mechanisms rely solely on pension misbeliefs and the potential for reform.
Online Appendix \zref{sec:ApenEmpiricalMen} documents similar employment and misbelief patterns for men, who were not subject to a reform, consistent with the view that this misbelief channel exists in the absence of a reform.

\section{Model \label{Model}}

Section \ref{sec:Rational-Expectation-Model} presents the baseline standard complete information model.
Section \ref{sec:Rational-Inattentive-Model} introduces two additions: objective uncertainty about government pension policy and costly information acquisition about this uncertain policy.

\subsection{Complete Information Baseline \label{sec:Rational-Expectation-Model}}

Key features are summarized before diving into details.
The model's decision-making unit is a household containing a couple or a single woman, but when a husband is present, his labor supply is inelastic.
The household maximizes lifetime utility from bequests, leisure, and equivalized consumption by choosing consumption, labor supply, and savings.
Households face risk over i) whether they get an employment offer, ii) the wage associated with any offer, and iii) mortality.
The households receive non-labor income from state and private pensions after the relevant eligibility age for each.

In more detail, households are divided into four types indexed by $k$, based on the high or low education status of the female and the presence or absence of a partner.
Periods are indexed by the age of the female ($t$).
Each period, households choose how much to consume ($c_{t}$), how much to invest in a risk-free asset ($a_{t}$) with return $r$, and, if not involuntarily unemployed, how much of the women's time endowment (normalized to 1) to devote to wage labor ($1-l_{t}$) (40, 20 or 0 hours per week) at a wage offer ($w_{t}$) that evolves stochastically.
Unemployment ($ue_{t}$), where $ue_{t}=0$ indicates employment (presence of a wage offer) and $ue_{t}=1$ unemployment (the absence), also evolves stochastically.
The partner's labor supply is inelastic, and so his behavior is treated as deterministic.
Although doing so abstracts from leisure complementarities and joint retirement, Online Appendix \zref{sec:ApenEmpiricalFactors} shows that the partner's labor supply is not significant in explaining labor supply responses to one's own SPA, which is the focus of this model.
The wife receives the state pension once she reaches the SPA ($SPA$), a parameter varied to mimic the UK reform, and a private pension once she reaches the type-specific eligibility age ($PPA^{(k)}$), which is a type-specific parameter that does not vary with $SPA$.
Both pensions, $S^{(k)}(.)$ the state pension and $P^{(k)}(.)$ the private pension, are treated as type-specific functions of average lifetime earning ($AIME_{t+1}=\frac{(1-l_{t})w_{t}+AIME_{t}t}{t+1}$) \footnote{This is average yearly earnings, to keep notation in line with the literature I use the abbreviation Average Indexed Monthly Earnings, which is the variable US Social Security depends on.}.
From age 60, the women face a probability of surviving the period ($s^{{k}}_{t}$).
Finally, households value bequests through a warm glow bequest function \citep{DeNardi2004}.
The full vector of model state is $X_{t}=(a_{t},w_{t},AIME_{t},ue_{t},t)$.

\paragraph*{Utility.}
The warm glow bequest motive creates a terminal condition ($T(a_{t})$) that occurs in a period with probability $1-s^{(k)}_{t-1}$:
\[
T(a_{t}) = \theta\frac{(a_{t}+K)^{\nu(1-\gamma)}}{1-\gamma}
\]
where $\theta$ determines the intensity of the bequest motive, and $K$ determines the curvature of the bequest function and hence the extent to which bequests are luxury goods.
The functional form around $a_{t}+K$ is the household's utility from consumption (see below), so the warm-glow bequest approximately captures the utility a descendant gains from these assets, and hence altruism as a motive, whilst keeping parameters to a minimum.

Whilst alive, a household of type $k$ has the following homothetic flow utility:
\[
u^{(k)}(c_{t},l_{t})=n^{(k)}\frac{((c_{t}/n^{(k)})^{\nu}l_{t}^{1-\nu})^{1-\gamma}}{1-\gamma}
\]
where $n^{(k)}$ is a consumption equivalence scale that takes the value 2 if the household is a couple and 1 otherwise.
In other words, utility takes an isoelastic form, with curvature $\gamma$, over a Cobb-Douglas aggregator of consumption and leisure, with consumption weight, $\nu$.

\paragraph*{Initial and terminal conditions.}
ELSA interviews people from 50, but the model starts with women aged 55 because this is the youngest age with significant numbers of SPA self-reports for multiple SPA-cohorts, thus allowing me to initialize state variables ($a_{t}$ and $AIME_{t}$ but in the model with incomplete information, also beliefs) from the empirical distributions for different SPA-cohorts.
At age 100, the woman dies with certainty.

\paragraph*{Labor market.}
The female log wage ($w_{t}$) is the sum of a type-specific deterministic component, quadratic in age, and a stochastic
component:
\begin{equation}
	\log(w_{t})=\delta_{k0}+\delta_{k1}t+\delta_{k2}t^{2}+\epsilon_{t}\label{eq:earning}
\end{equation}
where $\epsilon_{t}$ follows an AR1 process with persistence $\rho_{w}$ and normal innovation term with standard error $\sigma_{\epsilon}$, and has an initial distribution $\epsilon_{55}\sim N(0,\sigma_{\epsilon,55}^{2})$.
The quadratic form of the deterministic component of wages captures the observed hump-shaped profile and is common in the literature.

The unemployment status of the woman ($ue_{t}$) evolves according to a type-specific conditional Markov process.
From 80, the woman can no longer choose to work; this is to model some of the limitations imposed by declining health.
As spousal income results from the confluence of wages, mortality, and pension income, it follows a flexible polynomial in age:
\begin{equation*}
	\log(y^{(k)}(t))=\mu_{k0}+\mu_{k1}t+\mu_{k2}t^{2}+\mu_{k3}t^{3}+\mu_{k4}t^{4}
\end{equation*}
This specification averages out and abstracts away from both idiosyncratic spousal income and mortality risk.
In effect, the household dies when the woman dies, and the husband's mortality risk only turns up insofar as it affects average income, as if husbands were a pooled resource amongst married women of type $k$.
This allows me to ignore transitions between married and single, which, while important for the wider labor supply behavior of older individuals \citep[e.g.][]{Casanova2010}, are of secondary importance to employment responses to the SPA.
The function $y^{(k)}(t)$ amalgamates spousal labor and non-labor income, including pensions.
Both female wage and spousal income are post-tax.

\paragraph*{Social insurance.}

Unemployment status is considered verifiable, so only unemployed women ($ue_{t}=1$) can claim the unemployment benefit ($b$).

The wife receives the state pension as soon as she reaches the $SPA$, which abstracts away from the benefit-claiming decision.
This is done for two reasons, both touched upon earlier.
Firstly, over 85\% of people claim the State Pension at the SPA, so, in terms of accuracy, little is lost by this simplification.
Secondly, this small fraction deferring receipt occurs despite deferral having been actuarially advantageous during the period studied.
This presents another puzzle for standard models of complete information, which generally imply acceptance of actuarially advantageous offers.
In Online Appendix \zref{sec:Extension}, I provide preliminary evidence that costly attention to pension policy may also help explain this claiming puzzle, but in the main text, I abstract from it.
Abstracting from it here gives the baseline model a chance to solve the excess sensitivity puzzle.

Lifetime average earnings ($AIME_{t}$) evolve until the woman reaches the age she starts to receive her private pension ($PPA^{(k)}$), at which point it is frozen.
Both state and private pensions are quadratic in $AIME_{t}$ until they reach their maximum, at which point they are capped.
Until being capped, the pension functions have the following forms
\begin{equation*}
	S^{(k)}(AIME_{t})=sp_{k0}+sp_{k1}AIME_{t}-sp_{k2}AIME_{t}^{2}
\end{equation*}
\begin{equation*}
	P^{(k)}(AIME_{t})=pp_{k0}+pp_{k1}AIME_{t}-pp_{k2}AIME_{t}^{2}
\end{equation*}
These pension functions abstract from the details of state and private pension systems but capture some key incentives in a tractable form.
The state pension is a complex, path-dependent function shaped by past and current regulations \citep[see ][]{Bozio2010}.
This functional form captures the dependence of the state pension on working history without getting into these difficulties.
Being type-specific allows $S^{(k)}(.)$ to capture indirect influences of education and marital status on the state pension; for example, being a stay-at-home mum counted towards State Pension entitlement (after the enactment of a reform).
Every private pension scheme is different, but the dependence of $P^{(k)}(.)$ on $AIME_{t}$ reflects the dependence of most defined benefit schemes on lifetime earnings.
This functional form less accurately reflects the structure of defined contribution systems, which are essentially savings accounts. However, the model captures retirement savings through a risk-free asset and starts after the statutory defined contribution eligibility age, beyond which they can be accessed without penalty.

\paragraph*{Total deterministic income.}
Combining spousal income, benefits, and private and state pension benefits into a single deterministic income function yields:
\begin{equation*}
	\begin{split}
		Y^{(k)}(t,ue_{t},AIME_{t}) & = y^{(k)}(t)+b\mathbbm{1}[ue_{t}=1]+ \mathbbm{1}[t\ge SPA]S^{(k)}(AIME_{t})\\
		& +\mathbbm{1}[t\ge PPA^{(k)}]P^{(k)}(AIME_{t})
	\end{split}
\end{equation*}

\paragraph*{Household maximization problem.}
The Bellman equation for a household of type $k$ is:
\begin{equation*}
	V^{(k)}_{t}(X_{t})=\max_{c_{t},l_{t},a_{t+1}}\left[u^{(k)}(c_{t},l_{t})+\beta \left(s^{(k)}_{t}E[V^{(k)}_{t+1}(X_{t+1})|X_{t}]+(1-s^{(k)}_{t})T(a_{t+1})\right)\right]
\end{equation*}
subject to the following budget, borrowing, and labor supply constraints:
\begin{equation}
	c_{t}+(1+r)^{-1}a_{t+1} =a_{t}+w_{t}(1-l_{t})+Y^{(k)}(t,ue_{t},AIME_{t}), \label{eq:Budget}
\end{equation}
\noindent\begin{minipage}{0.4\textwidth}
	\begin{equation}
		a_{t+1}\ge0 \label{eq:Borrowing},
	\end{equation}
\end{minipage}
\begin{minipage}{0.19\textwidth}\centering
	\begin{equation*}
		\text{\&}
	\end{equation*}
\end{minipage}
\begin{minipage}{0.4\textwidth}
	\begin{equation}
		ue_{t}(1-l_{t})=0 \label{eq:LaborConstraint}.
	\end{equation}
\end{minipage}

\subsection{Two Additions: Policy Uncertainty and Costly Attention \label{sec:Rational-Inattentive-Model}}

This section adds two features to the complete information model.
Section \ref{sec:PU-Model} introduces objective policy uncertainty in the form of a stochastic SPA, reflecting potential SPA variation over the lifecycle caused by pension reform.
Section \ref{sec:RI-Model} adds costly attention to the stochastic SPA, in the form of disutility for more precise information, allowing the model to capture misbeliefs.
These additions are introduced independently, resulting in three model versions: the baseline from Section \ref{sec:Rational-Expectation-Model}, a version with policy uncertainty and informed households, and the full model with households who are rationally inattentive about the uncertain policy.

\subsubsection{Policy Uncertainty: the Stochastic SPA\label{sec:PU-Model}}
To capture the objective policy uncertainty resulting from the fact that governments can and, sometimes, do change pension policy, I make the SPA stochastic.

Although the SPA does change, introducing an important dimension of uncertainty, changes are not sufficiently frequent to estimate a flexible stochastic SPA process. For this reason, I impose a parsimonious functional form on the stochastic SPA:
\begin{equation}
SPA_{t+1}=\min(SPA_{t}+e_{t},\overline{SPA}) \label{lawMotionSPA}
\end{equation}
where $\ensuremath{e_{t}\in\{0,1\}}$ and $e_{t}\sim Bern(\rho)$. So each period, the SPA may stay the same or increase by one year, as the shock is Bernoulli, up to an upper limit of $\overline{SPA} = 67$.
This captures a key aspect of pension uncertainty, that in recent years governments have reformed pension ages upward but generally not downward, whilst maintaining a simple tractable form.
The lowest SPA, I consider possible is the pre-reform age of 60.
Hence, as the law-of-motion only allows for increases, $SPA_{t}$ is bounded below by $\underline{SPA} = 60$ and above by $\overline{SPA} = 67$.

In the model, the variable $SPA_{t}$ represents the current best available information about the age the woman will reach her SPA, and as such, the data analog is the SPA the government is currently announcing for the woman's cohort.
Only one SPA cohort is modeled at a time.
So there is no conflict in having a single variable $SPA_{t}$ whilst, in reality, at a given point in time, different birth cohorts have different government-announced SPAs.

\subsubsection{Costly Attention (Rational Inattention) \label{sec:RI-Model}}
The second addition is a cost of information acquisition about the stochastic SPA. 
This addition allows the model to reflect misbeliefs, and the modeling choices made in modeling costly information acquisition are motivated by the patterns seen in the belief data discussed in Section \ref{subsec:Second-and-Third} and Online Appendix \zref{sec:ApenEmpirical}. 
Firstly, we observed that although individuals are very mistaken about their own SPA, the vast majority are aware that a reform has taken place (Online Appendix \zref{sec:Other-Beliefs}).
Hence, to capture the pattern of misbeliefs, it is important to model the intensive margin of information acquisition rather than the extensive margin of awareness or unawareness alone.
Secondly, Section \ref{subsec:Second-and-Third} showed that selection into SPA knowledge was important for explaining the relationship between misbeliefs and the employment response to the SPA.
Hence, the process of information acquisition must, at least in part, be a choice that allows beliefs to respond to the incentives to be informed.  
The rational inattention paradigm fits these desiderata very well. 
It allows highly flexible information acquisition, capturing both intensive and extensive margin choices, including the extremes from zero to exhaustive information acquisition.
It is a paradigm in which the chosen information is an optimal response to incentives and constraints, but which also naturally captures the unpredictable, random nature of information arrival, since the information acquisition strategy is stochastic.
This section lays out the technical details of how introducing consistent, incentive-driven, Bayesian learning, in the form of rational inattention, changes the model.

\paragraph*{Directly observed vs learnable states.}

As the goal of this paper is to explain the employment response at SPA, the stochastic SPA ($SPA_{t}$) is the only state variable subject to a cost of information acquisition.
The $SPA_{t}$ is not directly observed by the household.
Instead, learning about its value requires paying a utility cost (defined below).
In contrast, other stochastic states such as wages ($w_t$) and unemployment status ($ue_t$) are observed directly. 
This can be understood as these variables being more salient.
To aid the exposition of information acquisition, these directly observable states are combined into a single vector of salient states $X_{t} = (a_{t}, w_{t}, AIME_{t}, ue_{t}, t)$ (the collection of variables in $X_{t}$ is unchanged from the baseline model, but no longer reflects all states since the hidden $SPA_{t}$ is excluded).
Additionally decisions are grouped into a single variable: $d_{t} = (c_{t}, l_{t}, a_{t+1})$. 

\paragraph*{Within period timing of learning.}

Since the household does not directly observe $SPA_{t}$, it is a hidden state.
It is still a state as it is payoff-relevant, but since the household does not observe it, it cannot enter the decision rule.
Instead their decision will depend on what they believe the SPA to be, introducing a new state variable: the household's belief distribution over possible SPA values, denoted $\underline{\pi_{t}} = (\pi(spa))_{spa = \underline{SPA}}^{\overline{SPA}} \in \Delta(8) \subseteq \mathbb{R}^8$.

Each period, the household chooses an information strategy that specifies how they will acquire information about $SPA_{t}$, then, using the information they receive, they make their labor supply and savings decisions.
That is, first, the household selects a signal distribution or information strategy ($\underline{f_{t}}[X_{t},\underline{\pi_{t}}](z \mid SPA_{t})$), then, conditional on the signal drawn from this distribution ($z_{t} \sim Z_{t}$), it chooses actions ($d_{t}[X_{t},\underline{\pi_{t}}](z_{t})$).
The household makes this choice having observed the other state variables, which is why the function chosen depends on $X_{t}$ and $\underline{\pi_{t}}$.
The household can choose as a signal any valid probability distribution without additional constraints, allowing the model to flexibly capture relationships between $SPA_{t}$ and beliefs.  
This flexibility can capture the partial informedness that we see in the data, but also allows for (in the guise of a degenerate signal) the idea of looking up and remembering your SPA.
Hence, it captures the extensive margin (e.g., looking up and remembering your SPA) as well as the intensive margin of learning (e.g., learning from friends) that leads to the partial informedness we see in the data.

Although this choice is unconstrained, households suffer a utility cost for choosing more informative signals (discussed below).
Due to this information-acquisition cost, households will generally choose noisy signals, not because being perfectly informed is ruled out, but because it is rarely optimal.
Hence, the modeling device naturally reflects both the responsiveness of information acquisition to incentives (one of the key desiderata) and the random component of learning that is outside the direct control of any individual.

Since the information strategy ($\underline{f_{t}}$) specifies how the household acquires information, not what information they acquire ($z$), the choice of information strategy can be thought of as the active choice of conditions under which to pay attention to the randomly arriving information from diverse sources outside their control, such as older siblings, colleagues, or the media. 
For example, the appearance of news stories about pension reforms is beyond an individual's control, but whether to continue reading past the headline is an active choice.
Hence, this modeling device reflects many features of the messy real-world learning process.

\paragraph*{Bayesian learning.}
Beliefs are updated using Bayes' rule (starting at an initial belief distribution ($\underline{\pi_{55}}$)). 
The posterior after observing signal $z_{t}$ is:
\begin{equation}
	Pr_{t}(spa \mid z_{t}) =
	\frac{f_{t}(z_{t} \mid spa)\pi_{t} (spa)}
	{Pr_{t}(z_{t})} = \frac{f_{t}(z_{t} \mid spa)\pi_{t} (spa)}
	{\textstyle \sum_{spa'=60}^{\overline{SPA}} f_{t}(z_{t} \mid spa')\pi_{t} (spa')}  \label{BayesUp}
\end{equation}
Next period's prior is then formed by applying the law of motion for $SPA_t$ (Equation \ref{lawMotionSPA}) to the posterior:
\begin{equation}
	\pi_{t+1}(spa)=
	\begin{cases}
	(1-\rho)Pr_{t}(spa \mid z_{t})+ \rho Pr_{t}(spa-1 \mid z_{t}) & spa <\overline{SPA} \\
	Pr_{t}(spa \mid z_{t})+ \rho Pr_{t}(spa-1 \mid z_{t}) & spa = \overline{SPA}
	\end{cases} \label{eq:BayesLawMot}.
\end{equation}

This assumes households know the law of motion of $SPA_{t}$, but not its current value. 
As mentioned in Section \ref{subsec:Second-and-Third}, we cannot separately identify mistakes about the law of motion from mistakes about the current value, and this paper loads mistakes onto the current value.
What is key is that information frictions generate SPA misbeliefs, not whether those misbeliefs load onto $SPA_{t}$ or its law of motion.
I conjecture that, qualitatively, the conclusions would be robust to making the law of motion the unknown object by having households learn about $\rho$.

\paragraph*{Entropy and mutual information.}

The cost of attention is modeled using tools from information theory.
Entropy (in the information-theoretic sense) measures uncertainty: it captures the minimal space needed to store or transmit the information in a random variable. 
Mutual information measures how much this uncertainty is reduced by learning another variable. 
These concepts will define the utility cost of information acquisition (below).
\begin{definition}[Entropy]
	Let $X \sim P_X(x)$. The entropy of $X$ is the expected value of $-\log (P_X(x))$:  
	\[ H(X) = \mathbb{E}_{X}[-\log(P_X(x))] \] 
\end{definition} 
\begin{definition}[Conditional entropy]
	The conditional entropy of $X$ given $Y$ is defined as:  
	\[ H(X \mid Y) = \mathbb{E}_Y[H(X \mid Y=y)] \]
\end{definition}
\begin{definition}[Mutual Information]
	Let $X$ and $Y$ be random variables. The mutual information between them is the expected reduction in uncertainty about $X$ from learning $Y$:  
	\[ I(X, Y) = H(X) - H(X \mid Y) \]
\end{definition}
In this model, the utility cost of attention is proportional to the mutual information between the belief over SPA and the chosen signal capturing information acquired during that period.
At the end of this section, I discuss reasons for this choice of functional form.

\paragraph*{Utility with attention costs.}
The household faces a trade-off between more informed decision-making and the cost of acquiring that information. 
On the one hand, learning about the SPA improves savings and labor supply choices. 
On the other, acquiring more precise information comes at a utility cost.
So, the value of information is the instrumental value of making better saving and labor supply choices, while its cost is a direct utility cost.
This trade-off is captured in the period utility function:
\begin{equation}
	u^{(k)}(d_{t},\underline{f_{t}},\underline{\pi_{t}})=
	n^{(k)}\frac{((c_{t}/n^{(k)})^\nu l_{t}^{1-\nu})^{1-\gamma}}{1-\gamma}-{\lambda I(\underline{f_{t}};\underline{\pi_{t}})}\label{eq:UtilRI}
\end{equation}
Here, $I(\underline{f_t}; \underline{\pi_t})$ is the mutual information between the belief $\underline{\pi_t}$ and the chosen signal $\underline{f_t}$, and $\lambda$ is the cost of attention.\footnote{This paper abstracts from preference parameters heterogeneity, and so in particular misses heterogeneity in information acquisition cost. Although more educated people almost certainly have lower costs of information acquisition, as shown in Online Appendix \zref{sec:ApenEmpiricalBeliefs}, they are actually more mistaken about their SPA. This indicates that heterogeneity in incentives to learn, which are captured by the model and are larger for the less educated, is more important in this context.} 
Expanding the mutual information and combining terms gives:
\[
I(\underline{f_{t}};\underline{\pi_{t}}) = \sum_{spa}\sum_{z}\pi_{t}(spa)f_{t}(z|spa)\log\!\frac{f_{t}(z|spa)}{\sum_{spa'}\pi_{t}(spa')f_{t}(z|spa')}
\]
%
What does this cost of information acquisition represent?
While your SPA is a single number freely available online, looking it up does not capture the full costs of learning it.
These should include information processing, storage, and recall costs, as well as straightforward hassle or time costs.
For illustration, the author has paid the hassle cost of looking up his SPA but not the cognitive cost of remembering it.
Hence, I would show up in survey data as having SPA misbeliefs, and I cannot use my SPA in decision-making.
Thus, the minimum data- and model-consistent conceptualization includes both cognitive and hassle costs.
Given that the time cost of looking up one's SPA can be measured in seconds and yet misbeliefs are prevalent in the data, it seems likely that the cognitive cost of information processing is more important than the time cost, consistent with the modeling assumption of a utility rather than a time cost.

Of course, the pension system is multifaceted, and people find many facets confusing, whilst the model concentrates all costs of information acquisition on tracking the SPA.
Thus, it is possible that the cost of information may also capture learning and the resolution of uncertainty about the pension policy more broadly.
An extension in Online Appendix \zref{sec:Extension} explores household learning about actuarial adjustment for deferred claiming.

\paragraph*{Revelation of uncertainty.}

In practice, SPA eligibility was communicated to individuals in the UK via letter shortly before SPA, and the claiming process involved a phone call that explicitly clarified the implications of claiming. 
The model reflects these institutional features by resolving uncertainty upon pension receipt.
Upon reaching $SPA_{t}$, the woman learns her true $SPA_{t}$ and starts receiving the state pension.
So, the household knows that if they do not receive the woman's state pension benefits, she is below her SPA.
This also ensures households never unknowingly exceed their budget.

\paragraph*{Dynamic programming problem.}
The household's state now includes both the belief over SPAs and the latent true SPA
$$(X_{t},SPA_{t},\underline{\pi_{t}})=(a_{t},w_{t},AIME_{t},ue_{t},t,SPA_{t},\underline{\pi_{t}}),$$
The Bellman equation becomes:
\begin{align}
	V^{(k)}_{t}(X_{t},SPA_{t},\underline{\pi_{t}}) = & \nonumber \\
	\max_{d_{t},\underline{f_{t}}}E\Big[u^{(k)}(d_{t},\underline{f_{t}},\underline{\pi_{t}}) &+\beta \big( s^{(k)}_{t} V^{(k)}_{t+1}(X_{t+1},SPA_{t+1},\underline{\pi_{t+1}})+(1-s^{(k)}_{t})T(a_{t+1})\big) \Big]  \label{BellmanRI}
\end{align}
subject to the same constraints in Equations \ref{eq:Budget} - \ref{eq:LaborConstraint} as the baseline model, and where now the utility function includes an information cost as per Equation \ref{eq:UtilRI}.
As previously stated, the choice objects ($\underline{f_{t}}[X_{t},\underline{\pi_{t}}](z \mid SPA)$ and $d_{t}[X_{t},\underline{\pi_{t}}](z_{t}) $), are restricted to be functions of the observed state, and the hidden state ($SPA_{t}$) appears in the value function only because expectations and continuation value are conditioned on it.

A challenge buried in this Bellman equation is the formation of next-period beliefs, which, due to Bayesian updating, depend upon the full distribution of the signal.
Hence, we need the solution to form the continuation value.
This problem is taken up in Section \ref{sec:Model-Solution}.

\paragraph*{Functional form of attention cost. \label{sec:Assumption}}

The information acquisition cost is key to the model mechanisms.
I assume it is proportional to the expected entropy reduction for three reasons.

Firstly, a cost of information acquisition that is directly proportional to mutual information is among the most common in the costly information literature, leading to two important advantages.
It is tractable as many useful results are available for this functional form,\footnote{Until \cite{miao2024dynamic} extended results from \cite{Steiner2017} to universally posterior separable functions, we only knew how to solve the dynamic rational inattention model with entropy-based cost of attention.} and it follows a convention.
Tractability is important in models of costly information, which can become too complex to solve, and following a convention has merit because it restricts the degrees of freedom available to fit the data.

Secondly, as argued by \cite{mackowiak2018rational}, this functional form offers a disciplined behavioral model by replicating numerous empirically supported departures from classical models.
It endogenously generates behaviors that look like heuristics, or rules-of-thumb, observed often enough to be christened biases in the behavioral literature.\footnote{For example, \cite{Koszegi2020} show this attention cost generates mental budgeting (quantity allocated to a category being fixed and composition changing) and naive diversification (composition being fixed and quantity allocated changing) in different situations.
	\cite{Caplin2019} show it leads to consideration sets.}

Thirdly, reasons exist to believe that the cost of cognition depends on entropy.
The information-theoretic concept of entropy sets a lower bound on efficient transmission and storage of information.
Thus, if the brain processes information efficiently, mutual information should factor into the ideal cost of attention function.
This is not to say that an ideal cost of attention function would be linear in mutual information, and recent works such as \cite{caplin2022rationally} generalize the traditional entropy penalty in multiple ways.
Laboratory evidence \citep[e.g.][]{dean2023experimental,bronchetti2023attention} indicates that the entropy-based cost of attention omits features of human attention that other cost functions better capture, although other experiments find it adequate \citep[e.g.][]{fuster2022expectations}.
Outside of such a controlled setting, however, it is not always clear which departures from the entropy-based costs are most relevant or whether sufficient data variation exists to identify their extra parameters.
As it seems that entropy enters an ideal cost function, my cost function can be considered a first-order approximation over this dimension.
        
\section{Model Solution\label{sec:Model-Solution}}

Solving a dynamic rational inattention model without suppressing the belief distribution as a state variable represents an important contribution of the paper.
To deal with the high-dimensional state $\underline{\pi_{t}}$ (beliefs) and a high-dimensional choice $\underline{f_{t}}$ (signal) introduced by rational inattention, I combine theoretical results into a solution method for dynamic rational inattention models with persistent beliefs, such as the one in this paper.
The solution method can be considered general-purpose since, one, it stores the belief distribution non-parametrically, and two, it does not impose restrictions on the data-generating process.
The only substantive restriction it imposes on the class of dynamic rational inattention models with this entropy-based cost of attention is that the problems must be discrete choice so that it falls within the class solved by \cite{Steiner2017}.
All computational methods require discretization, so this restriction can be seen as a computational approximation.
Due to this restriction, I discretize the assets and labor supply choices.
Section \ref{sec:Solution-RI} explains the general-purpose method, and Section \ref{sec:Solution_This_Model} details specific to solving the model of this paper.

\subsection{Solving Dynamic Costly Attention Models with History-dependent Beliefs \label{sec:Solution-RI}}
This section presents a solution method for dynamic rational inattention models with persistent beliefs.
I use the model of retirement decision from this paper to explain the method, but it applies to any dynamic, discrete-choice, rational-inattention model.
Section \ref{sec:SSM} outlines key results from \cite{Steiner2017}.
Section \ref{sec:SolMethod} uses these results and presents the method.

\subsubsection{Analytic Foundations of Solution Method \label{sec:SSM}}

\cite{Steiner2017} shows that a wide class of models has logit-like solutions.
Specifically, the class of dynamic discrete choice models subject to an additively separable entropy penalty for information acquisition about the state variable, of which the model presented in this paper is an example.  
The key results from their paper that are needed to understand the solution method are explained below.
The results are general, but I use my model to explain them to avoid extra notation.

If we define the effective conditional continuation value (i.e., the expected continuation value conditional on taking a given action) as:
\begin{align}
	\overline {V}^{(k)}_{t+1}(d_{t}, X_{t},SPA_{t},\underline{\pi_{t}})  = &  \nonumber \\
	E \big[ s^{(k)}_{t} V^{(k)}_{t+1}(X_{t+1},SPA_{t+1},\underline{\pi_{t+1}}(d_{t})) &+(1-s^{(k)}_{t})T(a_{t+1})\big| d_{t}, X_{t},SPA_{t},\underline{\pi_{t}}\big], \label{eq:effContVal}
\end{align}
where expectations are over $X_{t+1}$ and $SPA_{t+1}$ (Section \ref{sec:SolMethod} below describes finding $\underline{\pi_{t+1}}(d_{t})$), then the Bellman equation \ref{BellmanRI} becomes:
\[
V^{(k)}_{t}(X_{t},SPA_{t},\underline{\pi_{t}})=\max_{d_{t},\underline{f_{t}}}E\Big[u^{(k)}(d_{t},\underline{f_{t}},\underline{\pi_{t}})+\beta \overline{V}^{(k)}_{t+1}(d_{t}, X_{t},SPA_{t},\underline{\pi_{t}}) \Big] \label{BellmanRI2}.
\]
\cite{Steiner2017} show the optimal information acquisition strategy is to receive an action recommendation as signal, which results in a one-to-one mapping from signals to actions.
The intuition for this is that, since information is costly and the only benefit of information is making better choices, gathering any information not directly reflected in a choice would be wasteful. 
The linearity of information costs in mutual information is key to this one-to-one mapping; without it, it is possible that delaying or advancing information acquisition might decrease the cost of information.
Using this mapping, we can substitute actions for signals and the conditional choice probabilities ($d_{t}|SPA_{t} \sim \underline{p_{t}}(.|SPA_{t})$) for the signal function  ($\underline{f_{t}}$) throughout the problem, since one is simply a relabeling of the other.
Thus, we can combine the choice of a stochastic signal function ($\underline{f_{t}}$) and a deterministic decision conditional on the signal ($d_{t}(z_{t})$) into a single choice of a stochastic decision ($d_{t}|SPA_{t} \sim \underline{p_{t}}(.|SPA_{t})$).

With the problem now re-expressed in terms of choice probabilities, it is a calculus exercise to solve for the optimal distribution of actions, which they show has actions that are distributed with conditional choice probabilities $d_{t}|SPA_{t} \sim \underline{p_{t}}(.|SPA_{t})$ and associated unconditional probabilities $d_{t} \sim \underline{q_{t}}(.)$    (i.e., $q_{t}(d)= \sum_{spa=\underline{SPA}}^{\overline{SPA}} \pi(spa)p_{t}(d|spa) $) that satisfy:
\begin{equation}
	\textstyle p_{t}(d|spa) = \frac{\exp\left(n^{(k)}\frac{\left(\left(\frac{c}{n^{(k)}}\right)^\nu l^{1-\nu}\right)^{1-\gamma}}{\lambda(1-\gamma)}  +\log(q_{t}(d)) +\frac{\beta}{\lambda} \overline{V}^{(k)}_{t+1}(d, X_{t},SPA_{t},\underline{\pi_{t}})\right)}
		{ \sum_{d^{\prime}\in\mathcal{C}}\exp\left(n^{(k)}\frac{\left(\left(\frac{c^{\prime}}{n^{(k)}}\right)^{\nu} l^{\prime 1-\nu}\right)^{1-\gamma}}{\lambda(1-\gamma)}+\log(q_{t}(d^{\prime}))+\frac{\beta}{\lambda} \overline{V}^{(k)}_{t+1}(d^{\prime}, X_{t},SPA_{t},\underline{\pi_{t}})\right)} \label{eq:logit},
\end{equation}
\begin{equation}
	\resizebox{.9\hsize}{!}{$ \textstyle \underline{q_{t}} = \argmax_{\underline{q}} \sum_{spa} \pi_{t} (spa) \log \left( \sum_{d\in\mathcal{C}}q(d)\exp\left( n^{(k)}\frac{((c/n^{(k)})^\nu l^{1-\nu})^{1-\gamma}}{\lambda(1-\gamma)}+\frac{\beta}{\lambda} \overline{V}^{(k)}_{t+1}(d, X_{t},SPA_{t},\underline{\pi_{t}}) \right)\right) $} \label{eq:opq}.
\end{equation}
The logit-like structure captures how choice probabilities reflect not only utility differences but also information costs.
In contrast to the better-known random-utility multinomial logit, the source of stochasticity is noise in the information, which is not pushed to zero because it would be costly to do so.
This distinction is reflected by the presence of the endogenous objects $q_{t}(d)$ on the right-hand side of Equation \ref{eq:logit} and leads to substantive predictive differences between the two logit foundations discussed in \cite{Matejka2015}.
The similarity to EV1 RUM logit arises because the EV1 distribution maximizes entropy for a given expected utility, making it the least costly way to achieve that utility level \citep{Matejka2015}.

\subsubsection{General-Purpose Solution Method \label{sec:SolMethod}}  

At its core, the solution method is to solve Equation \ref{eq:opq} for $\underline{q_{t}}$ and substitute the solution into \ref{eq:logit} to get $\underline{p_{t}}$.
This basic description corresponds to an infeasible brute-force version of my solution method.
The problem with this brute-force version is that solving the high-dimensional optimization problem in Equation \ref{eq:opq} at every point in the state space, which contains the high-dimensional state $\underline{\pi_{t}}$, is computationally infeasible.
My solution method leverages the fact that there are actions that a rationally inattentive agent will never take for which there are conditions we can check (such as being strictly dominated) that are less computationally costly than the optimization in Equation \ref{eq:opq}.
My algorithm uses these conditions to filter out actions that will never be taken before solving Equation \ref{eq:opq}.
In my application, this filtering reduces the number of potential choices by two orders of magnitude and often finds a solution without solving the optimization problem. 
High-level pseudocode summarizing the algorithm, which may help follow the discussion that follows, is in Online Appendix \zref{sec:ApenComp}.
To get to a complete description of the algorithm, two hurdles must be passed.

The first hurdle is that knowing which belief next period will result from an action this period requires knowing the full probability distribution of actions.
This follows because we do not know how strong a signal an action is for a given SPA unless we know how likely households are to take that action, given other possible SPAs.
It follows that the conditional effective continuation value ($\overline{V}_{t+1}$) is not known, even though next period's value function ($V_{t+1}$) is known, because we do not know the beliefs tomorrow that will result from an action today ($\underline{\pi_{t+1}}(d_{t})$), which, as a state, enters $V_{t+1}$.
To see this, substitute the distributions of actions for the distribution of signals in the Bayesian updating formula \ref{BayesUp} and apply the results from Equations \ref{eq:logit} and \ref{eq:opq} to get:
\begin{align*}
	Pr(spa|d_{t}) = & \frac{ \textstyle \pi_{t} (spa) \exp\left(n^{(k)}\frac{((c/n^{(k)})^\nu l^{1-\nu})^{1-\gamma}}{\lambda(1-\gamma)}+\beta \overline{V}^{(k)}_{t+1}(d, X_{t},spa,\underline{\pi_{t}}))\right)}
	{\textstyle \sum_{d^{\prime}\in\mathcal{C}} q_{t}(d^{\prime}) \exp\left(n^{(k)}\frac{((c^{\prime }/n^{(k)})^{\nu} l^{\prime 1-\nu})^{1-\gamma}}{\lambda(1-\gamma)}+\beta \overline{V}^{(k)}_{t+1}(d^{\prime}, X_{t},spa,\underline{\pi_{t}}))\right)}.
\end{align*}
Then the prior at the start of next period ($\underline{\pi_{t+1}}$) is formed by applying the law of motion of $SPA_{t}$ (Equation \ref{lawMotionSPA}) to this posterior as per \ref{eq:BayesLawMot}. That is:
\begin{equation*}
	\pi_{t+1}(spa)=(1-\rho)Pr_{t}(spa|d_{t})+ \rho Pr_{t}(spa-1|d_{t}).
\end{equation*}
Thus, beliefs given choices ($\underline{\pi_{t+1}}(d_{t})$) are a function of the posterior, which depends not only on the exponentiated payoff but also on $\underline{q_{t}}$.
So, we need a solution ($\underline{q_{t}}$) to know $\underline{\pi_{t+1}}(d_{t})$ and hence to form the effective conditional continuation values (Equation \ref{eq:effContVal}).

\cite{Steiner2017} evade this difficulty by removing the beliefs from the state space and replacing them with the full history of actions.
They can do this because, given initial beliefs, the full history of signals, or equivalently actions, perfectly predicts the beliefs in period $t$.
This is an inspired step in their proof that extends \cite{Matejka2015} to the dynamic case, as it allows them to show we can ignore the dependence of continuation values on beliefs.
For applied structural modeling, it is often a non-starter because it introduces redundant information into the state space.
If two action histories lead to the same beliefs, they do not truly represent different states.\footnote{In \cite{Steiner2017}, past actions can affect beliefs and current utility.  Hence, two histories leading to the same belief might represent truly different states. This is not the case here. }
Redundant information in the state space is problematic, as the curse of dimensionality often makes this the binding constraint to producing richer models.
That the redundant information grows exponentially with the number of periods moves this from problematic to a non-starter for many applications.

Hence, I rely on the theoretical results of \cite{Steiner2017}, which used the history of action state-space representation, but in practice, I use the more compact belief state-space representation for the actual computational work.
To get around the issue that I need $\underline{q_{t}}$ to know $\overline{V}_{t+1}$ to work with the belief state-space representation, I use a simple guess-and-verify fixed-point strategy.
First, I guess a value $\tilde{\underline{q_{t}}}$ and use this to form next period beliefs conditional on having taking an action ($\underline{\pi_{t+1}}(d_{t})$) using Equations \ref{BayesUp} and \ref{eq:BayesLawMot}. 
With these, I form the effective conditional continuation value ($\overline{V}_{t+1}$) using Equation \ref{eq:effContVal}.
Then given $\overline{V}_{t+1}$ I solve \ref{eq:opq} for $\underline{q_{t}}$.
If the resulting $\underline{q_{t}}$ is close to $\tilde{\underline{q_{t}}}$, I accept this solution. Otherwise I set $\tilde{\underline{q_{t}}} = \underline{q_{t}}$ and repeat.\footnote{Although I have not proved this is a contraction mapping, the fixed point iteration always converges and generally in relatively few iterations.}

By increasing the computation required at each state, this solution to the first hurdle, however, exacerbates the second: the high computational demands resulting from the high-dimensional state $\underline{\pi_{t}}$.
Previously, models of dynamic rational inattention have generally avoided this problem by suppressing the belief distribution as a state variable \citep{miao2024dynamic,Armenter2019,turen2023state,Macaulay2021,Porcher2020}.\footnote{Sometimes this is justified as an explicit information sharing assumption in the model. Often, it is justified by noting that local posterior invariance \citep{caplin2022rationally} extends to global posterior invariance if all actions are taken with positive probability. However, \cite{Caplin2019} show that solutions are rarely strictly interior as rational inattention often implies consideration sets. Hence, extending local posterior invariance to a global property is restrictive.}
Although potentially reasonable in specific applications, suppressing beliefs prevents dynamic rational inattention from modeling situations in which beliefs matter and vary across individuals, as is the case for pension beliefs in the UK.
Heterogeneous subjective beliefs that impact choices have been documented in portfolio selection \citep{giglio2021five}, school choice \citep{kapor2020heterogeneous}, and compliance with health guidelines \citep{bhalotra2025maternal}. 
Thus, suppressing beliefs as a state variable imposes real limitations on the domain of applicability of rational inattention.
Moreover, these applications could potentially benefit from the dynamically consistent Bayesian model of incentive-driven belief formation offered by rational inattention.

My solution method keeps the belief distribution as a state whilst leveraging results of \cite{Caplin2019} to lighten the computational burden.
They show that often rational inattention implies consideration sets.
Hence, the solving conditional choice probabilities (CCPs) $\underline{p_{t}}$ are sparse.
That is, households take various actions with zero probability.
This sparsity reflects decision-makers' ability to rule out some actions as clearly not beneficial without further information.
To illustrate this idea, a household with total cash-on-hand of \pounds 10,000 can immediately rule out saving \pounds 9,999 without gathering any information about their SPA.

I propose two criteria that ex-ante identify actions that will be taken with zero probability, without solving the optimization problem.
I then remove these from the decision problem.
This filtering step always reduces the dimensionality of the optimization in Equation \ref{eq:opq}.
Moreover, if a single action remains after filtering, we have solved the problem without further calculation.
For my model, filtering leaves a single action in over 50\% of cases.

The first and simplest criterion for culling actions is removing strictly dominated alternatives.
The agent is rationally inattentive and so will never select an action strictly dominated in all possible realizations of the SPA.\footnote{Thinking about the wider applications of this solution method, rational inattention does not rule out choices that appear ex post strictly dominated, only those strictly dominated across all possible states of the world. For example, if a firm offers a multi-dimensional product like a mortgage that depends on the state of the economy, the rationally inattentive agent will never choose a product that is worse on all dimensions in all states of the world. They may choose a product that is ex post strictly dominated across all dimensions in the realized state of the economy because they assign some positive probability to the state being other than it actually is and believe that the company offers optimal products in that state. Hence, although choices appear strictly dominated ex post, they are not strictly dominated ex ante.}
Hence, all actions strictly dominated across all realizations of $SPA_{t}$ can be removed.
Checking this first criterion is helpful at two points in the procedure.
Firstly, before making an initial guess for $\tilde{\underline{q_{t}}}$, by removing any actions strictly dominated across all possible \textit{joint} combinations of $SPA_{t}$ realization and belief on the belief grid $\underline{\pi_{t+1}}$.
Doing this before entering the loop that solves for $\overline{V}_{t+1}$ reduces unnecessary computational burden in that fixed point iteration for $\underline{q_{t}}$.
However, it imposes a much stricter condition, dominant across all joint realizations of $SPA_{t}$ and $\underline{\pi_{t+1}}$, than needed to drop an action, dominant across all realizations of $SPA_{t}$.
Therefore, having made an initial guess for $\tilde{\underline{q_{t}}}$, and so having a prediction for next period's beliefs given any action ($\underline{\pi_{t+1}}(d_{t})$) and hence the conditional continuation value, I secondly remove actions strictly dominated across all realizations of $SPA_{t}$ for each belief in the belief grid as I loop over beliefs to find the policy rule at those beliefs.
I do this for each belief during each iteration of the loop that solves for $\overline{V}_{t+1}$.

In my model, the dimension reduction achieved by dropping strictly dominated actions is large, often two orders of magnitude.
Abstracting from borrowing constraints, the household faces 1,500 options, 500 saving levels, and 3 labor supply choices.
A household will never assign positive probability to more actions than the random variable they are learning about ($SPA_{t}$) has points of support.
$SPA_{t}$ has two points of support at the age of 65, increasing to 8 at age 59.
Hence, filtering often reduces the initial choice set in the high hundreds (once we impose borrowing constraints) to single digits.
The runtime required to perform a single filtering is negligible compared to the runtime required to solve Equation \ref{eq:opq}.

Removing strictly dominated actions only uses ordinal preferences.
The second criterion used to filter also uses the cardinal preferences encoded in expected utility.
It exploits the necessary and sufficient condition from \cite{Caplin2019}.
Using these, it is easily shown (see Online Appendix \zref{sec:Extending-CDL}) that if there exists a decision $d^{\star}=(c^{\star},l^{\star})$ which satisfies:
\begin{equation}
	\sum_{spa} \pi_{t}(spa) \frac{\textstyle \exp\left(n^{(k)}\frac{((c/n^{(k)})^\nu l^{1-\nu})^{1-\gamma}}{\lambda(1-\gamma)}+\frac{\beta}{\lambda} \overline{V}^{(k)}_{t+1}(d, X_{t},spa,\underline{\pi_{t}}))\right)}
	{ \textstyle \exp\left(n^{(k)}\frac{((c^{\star}/n^{(k)})^\nu l^{\star 1-\nu})^{1-\gamma}}{\lambda(1-\gamma)}+\frac{\beta}{\lambda} \overline{V}^{(k)}_{t+1}(d^{\star}, X_{t},spa,\underline{\pi_{t}}))\right)} < 1,
	\label{eq:CDL}
\end{equation}
for all other decisions $d=(c,l)$ then it is the only action taken ($q(d^{\star})=1$).
Unlike dropping strictly dominated alternative, which reduces the dimensionality, making solving Equation \ref{eq:opq} easier, checking Equation \ref{eq:CDL} is only beneficial when the optimal behavior is to take the same action in all realizations of $SPA_{t}$.
So, the benefits of checking condition \ref{eq:CDL} depend on how frequently, in the problem faced, it reveals the optimal choice without needing to solve an optimization.
When filtering does not leave a single action, I employ sequential quadratic programming to solve Equation \ref{eq:opq}, an algorithmic choice suggested by \cite{Armenter2019}.

Online Appendix \zref{sec:ApenComp} details two other computational difficulties.
Firstly, the large state space significantly increases the storage requirements for the solutions.
With this issue, the sparsity proved by \cite{Caplin2019} is again helpful as I can use sparse matrix storage techniques.
Secondly, when $\lambda$ is small, Equation \ref{eq:opq} can lead to underflow.

\subsection{Computational Details Specific to this Model \label{sec:Solution_This_Model}}

All versions of the model (the baseline, with policy uncertainty but informed households, and with rationally inattentive households) are solved by dynamic programming, specifically backward induction.
Beliefs ($\underline{\pi_{t}}$) and learning ($\underline{f_{t}}$) alter the nature of the within-period problem in the version with rationally inattentive households in some periods.
Only in some periods because $\underline{\pi_{t}}$ and $\underline{f_{t}}$  are only relevant before the SPA.
After the SPA, the true value is known, and so beliefs ($\underline{\pi_{t}}$) and learning ($\underline{f_{t}}$) about the SPA are irrelevant.
Periods after the SPA are solved, like periods in the other two versions, by simple search techniques to find the optimal choice from the discrete set of assets and labor supply choices.

In the version with rationally inattentive households, we proceed by backward induction from terminal age $t=100$ using standard techniques for the within-period problem until age $t=66$.
We can proceed back as far as age $t=67$ because $SPA_{t}$ is bounded above by 67, so the woman receives her state pension with certainty from that age onward.
Standard methods can also solve the period $t=66$ because, at this age, the household is perfectly informed.
Either she has reached her SPA and policy uncertainty has been resolved, or she infers $SPA_{t}=67$ with certainty, as she knows the data-generating process.
In this period, $\underline{\pi_{t}}$ is not a state variable, but $SPA_{t}$ is, as receipt of the state pension affects available resources.

At all earlier ages ($t<$66), if $SPA_{t} \le t$, then uncertainty has been resolved, meaning the model can still be solved using standard techniques, and, moreover, the exact value of $SPA_{t}$ is irrelevant. 
All that matters to the household is that they receive the benefit, so we can solve for a single representative SPA with $SPA_{t} \le t$.
Conversely, when the SPA is in the future ($SPA_{t} > t$),  the agent cannot infer the true value of the SPA, and so both the agent's beliefs ($\underline{\pi_{t}}$) and the true value of the SPA ($SPA_{t}$) are states and the agent needs to choose a learning strategy ($\underline{f_{t}}$).
Each year we proceed backward, the list of future potential SPAs ($SPA_{t} > t$) grows by one, increasing the combinations of  $\underline{\pi_{t}}$ and $SPA_{t}$  for which we need to solve a problem with uninformed learning agents that is not solvable by simple search techniques.
As $\underline{\pi_{t}}$ is a distribution over all future SPAs, its points of support also grow by one with each step in the backward induction.
For example, at age $t=65$, there are two potential future SPAs (66 and 67), and if $SPA_{t}$ takes on either of these values, the agent can no longer infer its true value, and so beliefs ($\underline{\pi_{t}}$) become a state and the choice of signal function relevant.
This growth of problem complexity along two related dimensions, rational-inattention-relevant potential future SPAs and the size of the belief distribution over them, continues until we reach $t=59$.
At this point, all SPAs 60-67 are future, and rational inattention is relevant regardless of the value of $SPA_{t}$ and the support of $\underline{\pi_{t}}$ is fixed.

\section{Estimation \label{sec:Estimation}}

The model is estimated by two-stage simulated method of moments.
The first stage estimates, outside the model, parameters of the exogenous driving processes and the initial distribution of state variables (a small number of parameters are also set, drawing on the literature).
Using the results from the first stage, the second stage estimates the remaining preference parameters $(\beta, \gamma, \nu, \theta, \lambda)$ using the simulated method of moments.

\subsection{First Stage \label{sec:First-Stage}}

The parameters of the wage process, the state and private pension system, and the unemployment transition matrix are estimated outside the model. The curvature of the warm-glow bequest and the interest rate are taken from the literature.

\paragraph*{Initial conditions.}
To set the initial conditions of the model, I need values for $a_{t},w_{t}$, $AIME_{t}$, $ue_{t}$, and in the version with rationally inattentive households $\underline{\pi_{t}}$.
Initial wages $w_{t}$ are drawn from the estimated initial wage distribution (see below), and all agents start as employed ($ue_{t}=0$).
Beliefs ($\underline{\pi_{t}}$) are initialized from the type- and SPA-cohort specific empirical distribution, and assets ($a_{t}$) and average earnings ($AIME_{t}$) from their joint type- and SPA-cohort specific empirical distribution.
The empirical counterpart used for assets is household non-housing non-business wealth.
Using the full work histories in the administrative data linked to wave 5 of ELSA, I construct a measure of $AIME_{t}$.
As this is only possible for a subsample, to estimate the joint distribution of $AIME_{t}$ and $a_{t}$, I impute missing $AIME_{t}$ values with a quintic in wealth and a rich set of observed characteristics (details in Online Appendix \zref{sec:ApenEstim}).
To initialize beliefs from point-estimate belief data, I assume that responses are draws from an individual's subjective belief distribution. 
This assumption is consistent with evidence from psychology that averaging multiple responses elicited from an individual improves accuracy \citep{vul2008measuring}. 
It also enables the construction of an individual's subjective belief distribution from point estimates. 
Given this assumption and imposing a homogeneous prior among individuals of a given type and cohort, the cross-sectional distribution will be the same as an individual's first period prior. 
Thus when simulating an individual of a given type and SPA cohort I set $\underline{\pi_{55}}$ equal to the empirical discrete distribution of self-reported SPAs at age 55.

\paragraph*{Wage equation.}

I assume wage data is contaminated with serially uncorrelated measurement error ($\mu_{j,t}$), leading to the following variant of Equation \ref{eq:earning} as data generation process:
\begin{equation*}
	\log(w_{j,t})=\delta_{k0}+\delta_{k1}t+\delta_{k2}t^{2}+\epsilon_{j,t}+\mu_{j,t}
\end{equation*}
for women $j$, of type $k$, and at age $t$.
The parameters of the age-dependent deterministic component of the wage process ($\delta_{k0},\delta_{k1},\delta_{k2}$) are estimated by type-specific regression.
The parameters of the stochastic component of the wage equation ($\rho_{w},\sigma_{\epsilon},\sigma_{{\epsilon,55}},\sigma_{\mu},$) are found minimizing the distance between the empirical covariance matrix of estimated residuals and the theoretical variance-covariance matrix of $\epsilon_{t}+\mu_{j,t}$ \citep[similar to][]{Low2010}.
I correct for selection into employment following \cite{French2005} by repeating the wage regression in the model, adjusting the model's wage equation parameters until the regression on the model and the data agree. 

\paragraph*{Pension systems.}
Both pensions are type-specific functions of average lifetime earnings.
These are estimated on the $AIME_{t}$ measures constructed from the administrative data described above.
As the state pension is relatively insensitive to education and the private pension to marital status, I simplify the state pension to be marital-status-specific and the private pension to be education-specific.
I estimate the private pension claiming age ($PPA^{(k)}$) as the type-specific mean earliest age women are observed with private pension income.

\paragraph*{Unemployment transition matrix.}
I classify a woman as unemployed if she claims an unemployment benefit and estimate type-specific transition probabilities in and out of unemployment.

\paragraph*{Stochastic State Pension age.}
I estimate the probability of an increase in the SPA, $\rho$, on the cumulative changes to the original female SPA of 60 experienced by reform-affected cohorts.
That is, I select the $\rho$ that minimizes the mean error in SPAs, given that the data-generating process is Equation \ref{lawMotionSPA}. 

\paragraph*{Parameters set outside the model.}
The curvature of the warm-glow bequest is taken from \cite{DeNardi2010} and the interest rate from \cite{ODea2018}.
Prices are deflated to 2013 values using the RPI.
Survival probabilities are taken from the UK Office for National Statistics life tables and combined with ELSA data to estimate type-specific survival probabilities following \cite{French2005}, details in Online Appendix \zref{sec:ApenEstim}.

\subsection{Second Stage \label{sec:Second-Stage}}

In the second step, moments are matched to estimate the preference parameters: the isoelastic curvature ($\gamma$), the consumption weight ($\nu$), the discount factor ($\beta$), and the bequest weight ($\theta$), as well as the cost of attention ($\lambda$) in the version with costly attention.

The 42 pre-reform moments of mean labor market participation and asset holdings from ages 55 to 70 were used to estimate $(\beta, \gamma, \nu, \theta,)$.
To avoid cohort effects or macroeconomic influences, a fixed-effect age regression was estimated, including birth-year effects, SPA-cohort-specific age effects, aggregate unemployment (to half a percentage point), and an indicator for being below the SPA.
Target profiles were then generated using these regressions with average pre-reform cohort values (details in Online Appendix \zref{sec:ApenEstim}).
In the model version with rationally inattentive households, $\lambda$ is identified from an additional moment: the reduction in the mean squared error of self-reported SPAs between ages 55 and 58 for the same pre-reform SPA-cohort as other targeted moments ($SPA=60$).
Since beliefs at 55 are initialized from the data (see Section \ref{sec:First-Stage}), the fit in that period is mechanical (a slight undershooting results from discretizing beliefs).
Hence, only beliefs at age 58 are targeted to identify $\lambda$, with beliefs at the two intervening ages (56 and 57) being untargeted.
As self-reported SPAs are assumed to reflect draws from a subjective belief distribution, I generate a random draw from each simulated individual's beliefs at age 58 ($\underline{\pi}_{58}$) and calculate the MSE of simulated self-reported SPAs using the distribution of these draws. 

\section{Results \label{sec:Results}}

Section \ref{sec:GoodFit} evaluates model fit and ability to replicate key facts on excess employment sensitivity, misbeliefs, and their relationship. 
Section \ref{sec:CompRefDep} compares these results to those of the reference-dependence preference retirement literature.
Section \ref{sec:Implications} explores the model implications.

\subsection{Model Evaluation \label{sec:GoodFit}}

This section presents the model fit and each version's ability to replicate the employment responses to the SPA and their relation to beliefs.
The objective probability of a pension reform occurring in any given year is estimated to be $\hat{\rho}= 0.102$. Other, less novel, first-stage parameter estimates are in Online Appendix \zref{sec:FirstEstim}.

\begin{figure}
	\flushleft
	\caption{Model Fit and Parameter Estimates}
	\begin{subfigure}{0.475\textwidth}
		\includegraphics[width=0.9\linewidth]{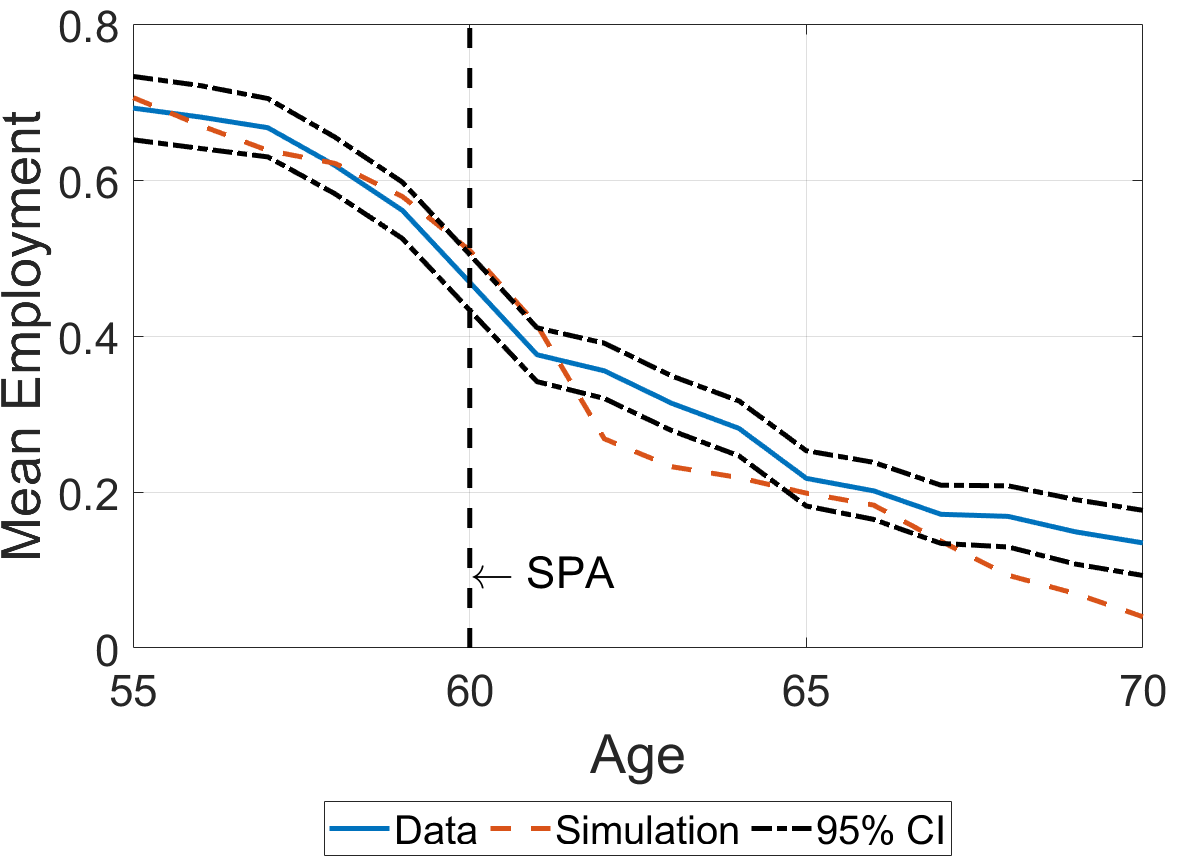}
		\caption{Employment Profile}
		\label{fig:Participation-Profile}
	\end{subfigure}%
	\begin{subfigure}{0.475\textwidth}
		\includegraphics[width=0.9\linewidth]{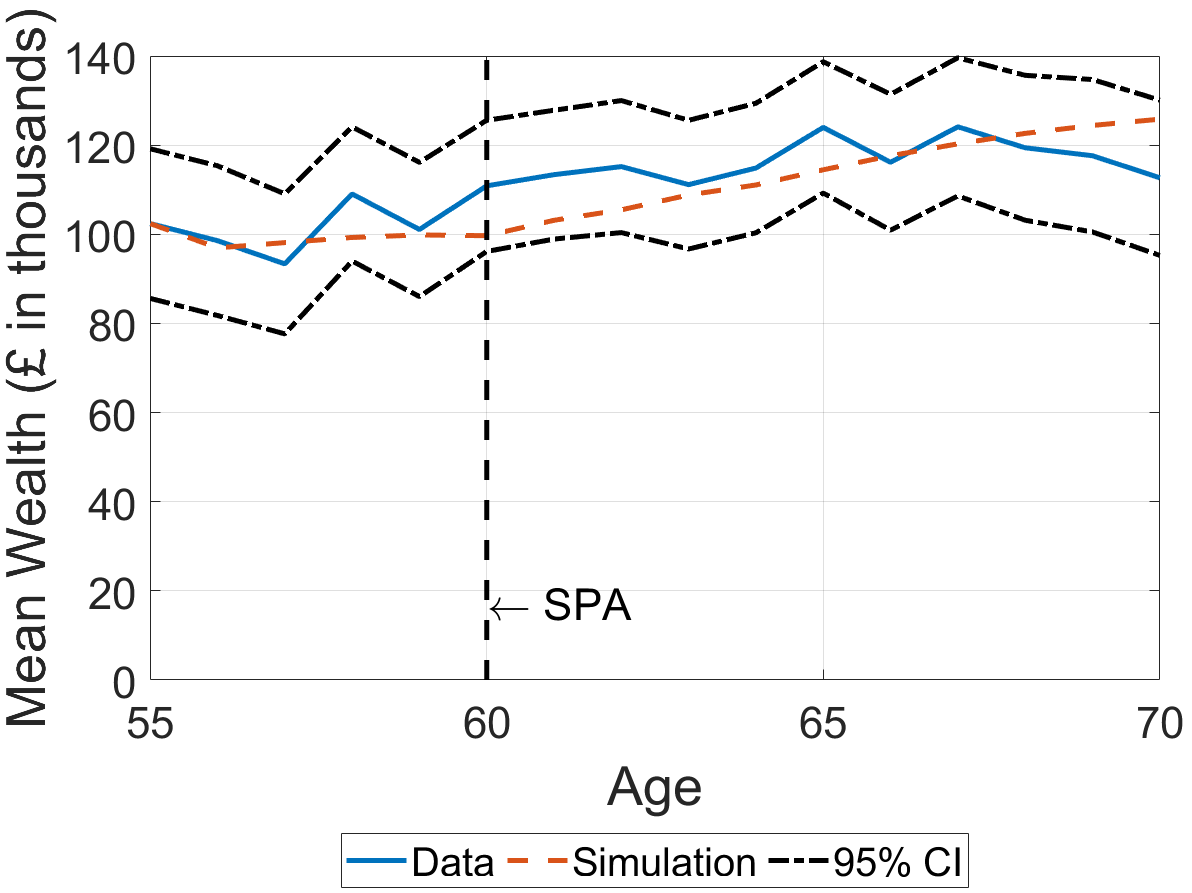}
		\caption{Asset Profile}
		\label{fig:Asset-Profile}
	\end{subfigure}
	
	\begin{subfigure}{0.475\textwidth}
		\includegraphics[width=0.85\linewidth]{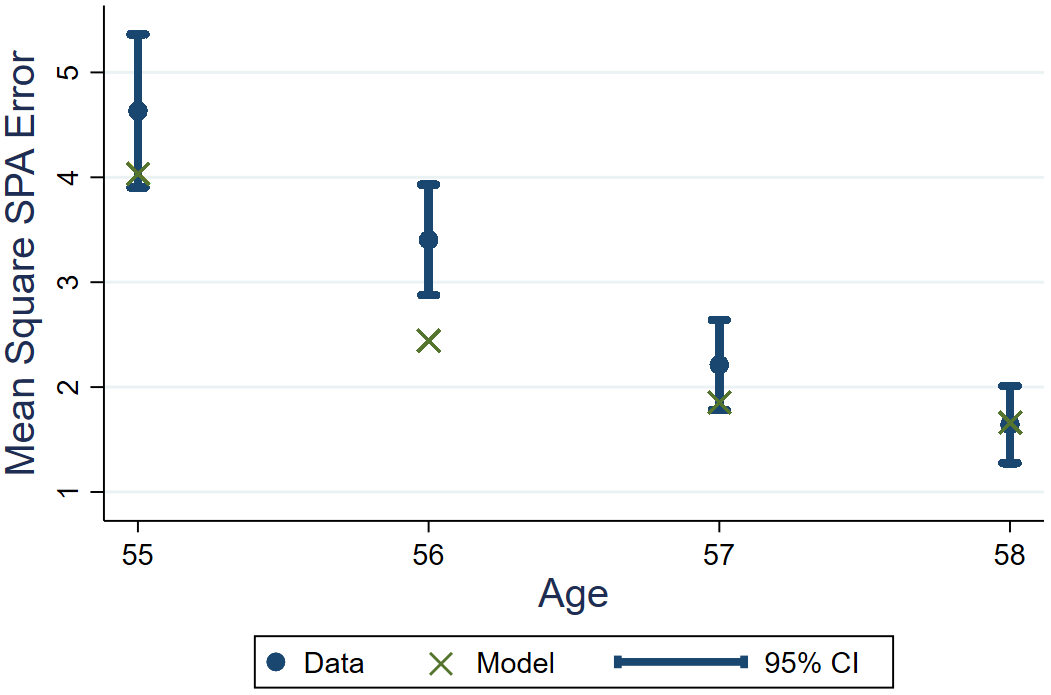}
		\caption{Belief Profile}
		\label{fig:Belief-Profile}
	\end{subfigure}
	\begin{subfigure}{0.475\textwidth}
		\resizebox{\columnwidth}{!}{
			\begin{tabular}[b]{ll}
				\multicolumn{2}{c}{\tablename } \\ 
				\hline
				$\nu$: Consumption Weight & 0.439 ( 0.0000038) \\
				$\beta$: Discount Factor & 0.985 ( 0.0000004) \\
				$\gamma$: Relative Risk Aversion & 3.291 ( 0.0000170 ) \\
				$\theta$: Warm Glow bequest Weight & 100 ( 2.626 ) \\
				$\lambda$: Cost of Attention & $6\times 10^{-8}$ ($ 2\times 10^{-11}$ ) \\
				\hline
				& \\
				& \\
			\end{tabular}
		}
		\caption{Parameter Estimates}
		\label{tab:estimates}
	\end{subfigure}
	
	\legend{ Panels (a)-(c) show model fit to targeted profiles, the empirical profile is for the pre-reform SPA cohort with a SPA of 60. Panel (d) shows estimated parameters (analytic standard errors in brackets calculated following \cite{newey1985generalized}).}
\end{figure}

Figures \ref{fig:Participation-Profile} and \ref{fig:Asset-Profile} show the fit of the version with policy uncertainty to the pre-reform employment and asset profiles when simulated with the pre-reform SPA of 60.
Since the baseline version and the version combining policy uncertainty with rational inattention produce similar fits to these static profiles, their profiles are placed in Online Appendix \zref{sec:AltMod}.
The mean square error of model-predicted and data beliefs used to identify $\lambda$ are presented in Figure \ref{fig:Belief-Profile}.
The value estimated of this parameter is $\hat{\lambda} = 6\times 10^{-8}$, and Table \ref{tab:estimates} reports all second-stage parameter estimates.\footnote {These are the estimated parameters obtained from estimating the version with rational inattentive households. To separate the impact of changes in other parameters from the introduction of rational inattention, I have also run an exercise in which I hold other parameters constant and estimate $\lambda$ from the additional moment. The results are available on request but change very little, as misbeliefs mostly strongly affect $\lambda$ anyway.}
Although the three model versions achieve very similar fits to the static employment profiles, the three versions predict very distinct responses to SPA changes.

To analyze this response to the SPA, I simulate the model with the SPAs observed in ELSA waves 1-7 ($SPA=60$, $SPA=61$, $SPA=62$) and repeat the regression from Section \ref{subsec:First-Puzzle} on the simulated data.
I adapt Equation \ref{eq:REg} to the model's simpler environment, estimating the treatment effect of being above the SPA on the hazard of exiting employment using a two-way fixed-effects difference-in-differences approach.
This regression includes the treatment indicator, full age, and cohort fixed effects (excluding period effects, which align with age in the model), and model counterparts to empirical controls (assets, marital status, and education).
As in Section \ref{subsec:First-Puzzle}, I repeat this on the subsample with above-median empirical assets (\pounds 28,500) before SPA.
Results are in Table \ref{tab:Reg-on-RE} 's top panel.
Column (5) repeats the empirical treatment effects from Columns (1) and (2) of Table \ref{tab:Treatment-Effect-different}.
The baseline model fails to match either.

This baseline's failure reflects the excess employment sensitivity puzzle, which prompted this investigation into policy uncertainty and costly attention.
To assess their impacts separately, I introduce them sequentially.
Column (2) shows that policy uncertainty alone has no effect.
This is because objective uncertainty is low (SPA changes are rare).
Both this version and the baseline fail to match the treatment effects for the whole population and for those with above-median assets at SPA, but are closer to the lower response of the richer subgroup.

\begin{table}
	\caption{Untargeted Model Fit to Regression Results}
	\label{tab:Reg-on-RE}
	\centering{}
	\resizebox{\columnwidth}{!}{ \begin{tabular}{rc c c c l}
			\hline
			& (1) & (2) & (3) & (4) & (5) \\
			& Baseline & Policy Uncert. & $\hat{\lambda} = 6\times 10^{-8} $ & $\lambda = 5.0\times 10^{-7}$  & Data (95\% C.I) \\
			\hline
			& \multicolumn{4}{c}{Treatment Effect being above SPA on the Hazard of Exiting Employment} & \\
			\hline
			Whole Population & 0.019 & 0.014 & 0.041 & 0.068 & 0.123 \textit{(0.0766,0.1696)} \\
			Assets \textgreater Median(\pounds 28,500) & 0.018 & 0.014 & 0.054 & 0.088 & 0.088 \textit{(0.0230,0.152)} \\
			\hline
			& \multicolumn{4}{c}{Treatment Effect Heterogeneity by Absolute SPA Error} & \\
			\hline
			Interaction & \textemdash & \textemdash & -0.047 & -0.046 & -0.082 \textit{(-0.1577,-0.0061)} \\
			\hline
			& \multicolumn{4}{c}{Treatment Effect Heterogeneity by SPA Error Positivity} &\\
			\hline
			Interaction & \textemdash & \textemdash & -0.047 & -0.046 & -0.207 \textit{(-0.3973,-0.0168)} \\
			\hline
	\end{tabular}}
	\legend{The top panel shows employment response across the wealth distribution (Table \ref{tab:Reg-on-RE}).
		The second panel shows heterogeneity in SPA labor supply response by the absolute size of self-reported SPA error at 58.
		The third panel shows heterogeneity in SPA labor supply responses by direction of self-reported SPA error at 58, and the third panel by the absolute size of the error.
		Results in Column (1) are from the Baseline, from Column (2) with policy uncertainty alone, Columns (3) and (4) from the full model with policy uncertainty and costly information, and Column (5) shows data counterparts. 
		There is no version without policy uncertainty, but with costly attention, since costly attention needs policy uncertainty to operate.
		Some results are identical to three decimal places but differ to four decimal places.}
\end{table}

Column (3) of Table \ref{tab:Reg-on-RE} adds households that are rationally inattentive about the policy uncertainty introduced in Column (2).
It shows that this model version matches the employment response to the SPA significantly better than the baseline or the policy-uncertainty versions, but still falls short of the data.
Costly attention closes 21\% of the gap for the whole population and 51\% for the richer subgroup, with only the richer subgroup's estimate falling within the 95\% confidence interval.
As explained in Section \ref{subsec:First-Puzzle}, the response among the wealthy subgroup is the ex-ante puzzle.
I therefore interpret a model’s ability to explain the response among those with above-median assets at SPA as its ability to explain the puzzle, even though the models struggle most with the aggregate response at the estimated parameters. 
It is worth noting that directly targeting the two treatment effects lets the baseline match the aggregate response but not the wealthy response (results available on request).

So, the introduction of rational inattention improves the model's ability to account for employment responses to the SPA in an economically significant way. 
This improvement is achieved without directly targeting these responses or introducing free parameters, as data-driven restrictions are imposed on the added model features: policy uncertainty is estimated from past SPA reforms and information friction from misbeliefs.

Is the increase in employment response from introducing rational inattention reasonable?
The introduction of rational inattention increases the impact of the SPA on the probability of exiting employment by 2.7 percentage points, and this increase is driven by two potential mechanisms: precautionary labor supply and wealth effects triggered by the resolution of pension uncertainty.
Upon reaching SPA, households in the model overestimate their own SPA by 0.7 years, on average, implying that the wealth shock they receive is roughly equivalent to learning they have an extra year of pension income.
More precisely since average yearly state pension benefits are \pounds 4,691, the implied wealth shock is \pounds 3,284.
Then assuming an otherwise constant hazard, the 2.7 percentage point increase in hazard implies an expected decrease in employment of 0.18 years of employment.
If this increase in hazard is driven purely by wealth effect, this implies, at the mean, a marginal propensity to earn (MPE) of -0.64.
Using employment responses to pension-wealth increases among women in Germany, \cite{artmann2023forward} estimate an MPE of -0.3.
Exploiting lottery winnings, \cite{imbens2001estimating} estimate an MPE of -0.11 in Massachusetts; \cite{cesarini2017effect} find MPEs in Sweden, in the range -0.17 to -0.15; and \cite{golosov2024americans} estimate an MPE of -0.5 in the United States.
Using changes in survivors' benefits, \cite{coyne2024household}, \cite{rabate2024labour}, and \cite{giupponi2019income} estimate MPEs of -0.3, -0.2, and -0.8, respectively.
Hence, the implied wealth effect would fall within the range estimated in the literature.

In addition to the wealth effect, the increase in employment could also be driven by precautionary motives.
Precautionary labor supply has proven more difficult to estimate than wealth effects, and, to the best of my knowledge, no estimates specifically related to pension uncertainty exist.
\cite{pistaferri2003anticipated} estimates precautionary labor supply and finds it to be small but significant, and \cite{jessen2018important} quantify precautionary labor supply motives as accounting for 2.8\% of hours worked among married men.
These estimates are less comparable to the model output than those of wealth effects, but perhaps a 2.7 percentage-point increase in the probability of exiting work can be considered of the same order as 2.8\% of hours.

It is important, however, to test model mechanisms as well as estimate them, as emphasized by \cite{fang2007testing}, among others.
Section \ref{subsec:Second-and-Third} discussed key patterns in the belief data, including the declining pattern of misbeliefs used to identify $\lambda$, and, crucially here, two patterns related to the relationship between misbeliefs earlier in life and employment responses later. 
These two patterns provide natural, untargeted moments for testing the model's mechanisms.

The first pattern was that individuals who are worse informed about their SPA in their late 50s exhibit smaller labor supply responses at SPA in their 60s as captured by the negative interaction term in Column (5) of Table \ref{tab:Treatment-Effect-different}.
Two opposing forces in the model link the accuracy of earlier SPA knowledge to labor supply responses to it.
Endogenous SPA knowledge implies that those less dependent on the SPA acquire less information about it.
Conversely, households that are less well informed due to unfavorable signal realizations rather than their choice of signal distribution face a larger shock upon learning their SPA, prompting a greater reaction.
Which of these two mechanisms dominates, and hence whether the model generates a positive or negative relationship between misbeliefs and employment response to the SPA, is not predetermined.
The middle panel of Table \ref{tab:Reg-on-RE} shows that the model generates a negative relationship, indicating the model reproduces the observed direction of this relationship, with the empirical counterpart repeated in the final column.

The second pattern was that those who over-estimate their SPA, as opposed to those who under-estimate, have a larger employment response upon reaching SPA as captured by the negative interaction term in Column (6) of Table \ref{tab:Treatment-Effect-different}.
This is because, all else equal, they receive a larger shock upon reaching their SPA.
The bottom panel also shows that the model replicates the direction of the dependence between SPA employment responses and SPA misbeliefs.

\subsection{Comparison to Reference point retirement \label{sec:CompRefDep}}

A prominent explanation for the sharp employment response at pension eligibility is reference-dependent preferences or norms that shift in utility from leisure at a pension age. 
This mechanism provides a compelling account of behavioral discontinuities and has been influential in explaining responses to statutory retirement ages \citep[e.g.][]{Seibold2021}.
However, reference dependence typically abstracts from widespread misbeliefs about the eligibility age.
Although reference dependence does not require complete information, the widespread misbeliefs documented here and in \cite{caplin2022rationally} about the age it posits as a salient reference point are somewhat in tension with it as an explanation.

More importantly, there is an advantage to having a more parsimonious explanation that accounts for both misbeliefs and excess sensitivity simultaneously. 
Table \ref{tab:QualPredict} summarizes the data patterns that this paper has tried to match and then indicates whether this pattern is qualitatively consistent with three different drivers of employment response: liquidity constraints, reference dependent preference, and rational inattention.
Of course, this is a slightly loaded comparison since reference dependence was not designed to explain misbeliefs, but it brings home the benefit of costly attention in parsimoniously explaining the joint distribution between beliefs and actions with one new parameter.
Moreover, the predictive value of misbeliefs for employment responses suggests they are related phenomena, something that follows naturally from the model of costly attention but fits uneasily with explanations that abstract from misbeliefs, such as reference-dependence. 

Studies in the reference-dependence literature generally introduce a reference point or utility kink to directly match the response at pension age.
Instead, I estimate an attention cost from belief data without directly targeting the puzzle. 
When I also choose a parameter value to directly target the puzzling response of the wealthy, in Column (4) of Table \ref{tab:Reg-on-RE}, costly attention accounts for 47\% of the gap for the whole population and completely accounts for the richer subgroup, with both estimates falling within the 95\% confidence intervals.
So, on this metric, costly attention explains excess employment sensitivity as well as reference dependence.

Directly targeting the employment response at SPA sacrifices some of the most appealing features of costly attention. 
Namely, it provides a unified framework to explain both the observed employment response and belief inaccuracy, and it allows us to bring in additional data on beliefs to identify the underlying mechanism. 
When the model is thus constrained to match observed misbeliefs, however, it only partially explains the excess employment sensitivity.

Therefore, it is worth considering why the model underestimates employment responses to the SPA rather than just pivoting to targeting the puzzle itself.
I consider two potential explanations.
Firstly, the model attributes all policy learning to the SPA, while real-world pension systems involve multiple complex features—some of which may also become salient around eligibility. 
This could understate the degree of learning at SPA and, hence, the size of the shocks received. 
Online Appendix \zref{sec:Extension} explores this possibility by extending the model to include uncertainty over deferral rules, though a lack of data on deferral rate beliefs makes this exercise more speculative.
Secondly, belief-driven mechanisms may operate in parallel with behavioral ones, such as reference dependence. 
There is suggestive evidence that framing effects, such as labeling eligibility as a retirement norm, can influence labor supply responses—consistent with reference-point models. 
I discuss this further in Online Appendix \zref{sec:RefDep}.
Online Appendix \zref{sec:Passive} also presents results from an estimated model version that includes, in addition to costly attention, a share of passive decision-makers \citep[as in][]{chetty2014active}, who retire mechanically at their SPA.
This extension provides a simple way to capture potential reference dependence, allowing for the possibility that both mechanisms are working in parallel.
This exercise indicates that reference dependence can explain the portion of the employment response not explained by misbeliefs.

Thus, the costly attention model is perhaps best understood not as a rejection of reference dependence, but as a complementary explanation that addresses dimensions (such as belief heterogeneity) not captured by reference-based preferences. 
In doing so, it might help explain why responses vary across individuals, cohorts, and institutional contexts.
For example, \cite{deshpande2024sticky} find smaller employment responses to the U.S. full retirement age during reform periods. 
If behavior were driven entirely by fixed preferences, such variation would be hard to reconcile. 
With costly attention, however, policy salience and information campaigns during reform periods can reduce misbeliefs, attenuating responses.
Further research could shed light on the relative importance of these two explanations.

\begin{table}[htbp] 
	\centering
	\caption{\label{tab:QualPredict} Qualitative predictions of each theory}
	\begin{tabular}{lcccc}
		\hline
		\textbf{Pattern in the data} & \textbf{Liquidity} & \textbf{Reference Dependence} & \textbf{Costly attention} \\
		\hline
		Drop concentrated at SPA                 & \checkmark & \checkmark & \checkmark \\
		Drop large above-median wealth            & $\times$ &  \checkmark & \checkmark \\
		People display SPA misbeliefs            & $\times$ &  $\times$ & \checkmark \\
		Misbeliefs decline as people age            & $\times$ &  $\times$ & \checkmark \\
		Smaller response if more mistaken at 58    & $\times$ & $\times$ & \checkmark \\
		Over-estimators react more strongly        & $\times$ & $\times$  & \checkmark \\
		\hline
	\end{tabular}
		\legend{This table simply recaps which of the theories discussed are qualitatively consistent with the data patterns investigated.}
\end{table}

\subsection{Model Implications and Predictions \label{sec:Implications}}

\paragraph*{Attention cost size.}
$\lambda$ is hard to interpret, having natural units of utils per bit.
While utils are known to be non-interpretable, denoting in bits is also problematic, as it exaggerates costs, since models contain far fewer learnable bits than reality.
Most models contain only single or double-digit bits of information, fewer than in an average sentence.
Reality holds vastly more information, making per-bit information cost a larger share of total model information.
To address both issues, I calculate the compensating asset that raises household utility as much as perfect SPA knowledge, effectively their willingness to pay to learn their SPA.
For $\hat{\lambda} = 6\times 10^{-8}$, compensating assets range from \pounds 6 at the 25th percentile to \pounds 14 at the 75th, with a mean of \pounds 11.
For $\lambda = 5\times 10^{-7}$, the mean is \pounds 31 (summary of compensating assets distributions for both $\lambda$ values in Online Appendix Table \zref{tab:Heto}).

The model finds such modest friction because the benefit from learning about your SPA is not of the order of the State pension benefit but of the difference this information makes in saving and labor supply decisions, and overall, these decisions are not that sensitive to this information.
These results are similar in magnitude to findings from other settings, both in terms of the implied sensitivity of decisions to information about pensions and the estimated welfare effects.
\cite{mastrobuoni2011role,dolls2018retirement,liebman2015would} find modest impacts of information about pension on choices. 
\cite{Luttmer2018} estimates that people would be prepared to part with 6\% of their pension benefit to remove all pension uncertainty.
Although not directly comparable to the estimates here since they estimate willingness to pay to remove \textit{all} policy uncertainty, and I estimate willingness to pay to be perfectly informed of \textit{current policy}, at the mean 6\% of the pension benefit in my setting is \pounds 180, which is not of too dissimilar a magnitude to the estimate presented above.

\paragraph*{The employment response to pension age reforms.\label{sec:oldAgeEmp}}
Rising old-age dependency ratios make increasing older individuals' employment a global policy priority, with pension ages seen as a key tool \citep[e.g.][]{landais2021retirement}.
This paper shows that misbeliefs from costly attention amplify employment responses at the SPA. 
This may naturally raise the question of whether misinformation makes the SPA a more effective tool to increase old-age employment.
In fact, the model generally implies it does the opposite.

\begin{table}
	\caption{\label{tab:AdEmp} Impacts of Reforming SPA with Informed and Uninformed Households }
	\begin{center}
		\begin{tabular}{ c  c  c  c   c  c   c   }
			\cline{1-6}
			SPA increased  & (1) - Informed &  (2)  - Uninformed  & \multicolumn{1}{|c}{ (3) } & (4) & (5) \\
			from 60 to: & Added Employment & Added Employment & \multicolumn{1}{|c}{ MC }  & WTP & MR \\
			\hline
			61 &	0.07 & 0.06  & \multicolumn{1}{|c}{ \pounds 3.50 }& \pounds 4.22 & \pounds 28.45 \\
			62 &	0.14 & 0.14  &\multicolumn{1}{|c}{  \pounds 4.00 }& \pounds 2.37 & \pounds 11.78 \\
			63 &	0.18 & 0.16  &\multicolumn{1}{|c}{  \pounds 4.50 }& \pounds 18.34 & \pounds 19.91 \\
			64 &	0.22 & 0.20  &\multicolumn{1}{|c}{  \pounds 5.00 }& \pounds 31.64 & \pounds 4.31 \\
			65 &	0.31 & 0.27  &\multicolumn{1}{|c}{  \pounds 5.50 }& \pounds 44.41 & \pounds 68.52 \\
			\hline
		\end{tabular}
	\end{center}
	\legend{ Employment increases over 56-65 from raising SPA from 60 to the age in Column (2) with costly attention and in Column (1) without it. Columns 3-5 show the financial impacts of an accompanying information letter campaign that moves people from uninformed to informed. Column (3) shows the marginal cost, Column (4) the model implied mean willingness to pay, and Column (5) the model-implied the marginal revenue.}
\end{table}

Column (2) of Table \ref{tab:AdEmp} shows the additional mean employment between ages 55-65 when the SPA is reformed from 60 to the value between 61 and 65 indicated for that row, in the model with $\hat{\lambda} = 6\times 10^{-8}$ where prior beliefs and other state variables are initialized to the values of the SPA 60 cohort.
Thus, this captures the response to an unanticipated SPA increase at age 55 from 60 to a later age, since households have savings, accrued pension entitlements, and the beliefs of a cohort that had a SPA of 60.
Column (1) shows results from the model with policy uncertainty but no attention costs.
Both versions show modest employment gains, but the increases are generally larger under costly attention. 
For example, for post-reform SPA 65, mean employment rises 0.27 years with attention costs vs. 0.31 without.
So, employment rises by up to 15\% more for informed households, which may seem at odds with the finding that costly attention increases employment responses at the SPA.

\begin{figure}
\caption{Additional Employment resulting from Increasing the SPA from 60 to 62 \label{fig:lpredict2}}
\center{}\includegraphics[scale=0.18]{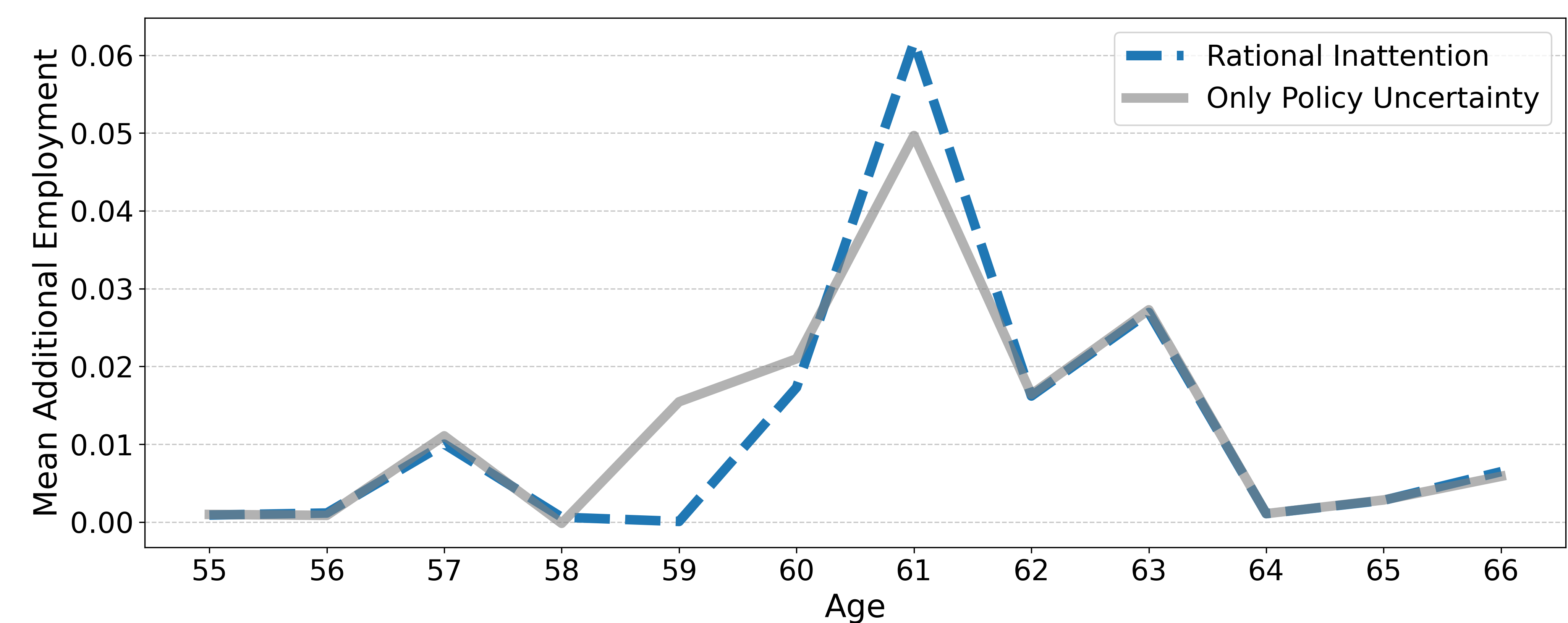}
\legend{For the two versions, employment increases resulting from a reform of the female SPA from 60 to 62.}
\end{figure}

This tension resolves when noting that rationally inattentive households respond less immediately to SPA increases.
Fully informed households internalize the change early, increasing work in their 50s.
Inattentive households react later when they realize their SPA has increased and they must compensate for lost earnings.
This compensatory effort reduces the difference that has already arisen but generally does not eliminate it, due to imperfect intertemporal substitution and lower employment at older ages.
This effect also inflates employment just before SPA, amplifying the drop at SPA.
Thus, costly attention yields smaller overall employment gains but a larger response at SPA, with this bunching driven by intertemporal shifts.
This may help us understand how small information frictions can generate relatively large employment effects, because much of the concentrated employment response reflects an intertemporal shifting of employment, and the welfare implications are not as high as they seem when we focus on the local effect.
This echoes other results in the literature, such as \cite{chetty2012bounds} and \cite{choukhmane_default_2025}, which find that apparently large deviations from optimizing behavior are explained by modest friction once we account for dynamics.
Figure \ref{fig:lpredict2} illustrates the intertemporal shifting of employment for an SPA rise to 62, for which the two effects roughly cancel out.

\paragraph*{The impact of information on response to pension age reforms.}
Columns (1) and (2) of Table \ref{tab:AdEmp} show added employment from an unanticipated SPA increase at age 55 in models with and without costly information.
The only difference is in Column (1), households know the SPA, and in Column (2), they do not.
Thus, the gap reflects the maximum potential impact of an annual information letter campaign.
Columns (3)-(5) assess such a campaign.

Column (3) reports the marginal cost of the information letter campaign.
After covering fixed costs, the only marginal cost is postage at \pounds 0.50/year (2013 prices, like the model).
Column (4) shows the willingness to pay (WTP) for the information campaign under each post-reform SPA.
Two forces drive WTP: higher SPAs reduce lifetime wealth (lowering WTP), but also, as it moves further from the pre-reform SPA of 60, the value of information rises.
Initially, the first effect dominates, reducing WTP.
From SPA 63 onward, the second dominates, and WTP increases.
Comparing Columns (3) and (4) shows WTP for information exceeds the campaign's marginal cost for all post-reform SPAs except 62.
For these reforms, the information campaign improves net welfare without accounting for added government revenue, but since the campaign also raises employment (see the difference between Columns (1) and (2)), the campaign is revenue-positive as quantified in Column (5), which presents marginal revenue from the campaign.
Though modest (because 1950s-born women had low earnings), marginal revenue exceeds marginal cost for all SPA reforms except 64.
Combining household and government gains, Columns (3)-(5) show the information campaign consistently raises total welfare, with benefits exceeding costs by 3.5 to 20.5 times.
Though absolute gains are modest, the experiment underscores a key point: informing individuals not only improves their welfare but also improves their policy responsiveness.
Additionally, as pointed out by \cite{dolls2018retirement}, the gains from information letter campaigns are not so modest when benchmarked against other policies that aim to increase old-age employment or retirement saving.

\section{Conclusion \label{sec:Conclusion}}

Mistaken beliefs are common, but their economic impacts are still not well-understood.
Using UK data, this paper shows that incorporating costly attention, which endogenously generates misbeliefs, into a retirement model explains both observed misbeliefs and the sensitivity of employment to pension eligibility ages.
Costly attention accounts for 51\% of the employment response gap between the model and the data when calibrated to observed beliefs, and all of the gap when not so constrained.
Given that both pension misbeliefs and excessive employment responses are across-country regularities, these insights may be cross-nationally relevant.

Endogenous information acquisition is key to explaining retirement behavior, but leads to the prior belief becoming a state variable.
This high-dimensional state variable significantly increases computational demands.
I propose a method for solving dynamic rational inattention models without suppressing beliefs as a state variable.
From the belief data, I estimate the mean willingness to pay to learn the SPA as \pounds 11.
Though small, this far exceeds the marginal cost of information letters.
Policy experiments show that, after most SPA reforms, households' willingness to pay for such letters exceeds their cost, but also that sending letters increases employment by up to 15\%.
Hence, the campaign raises additional tax revenue, which, for most SPA reforms, also exceeds the cost.
Considering total benefits to government and households, the campaign always improves welfare, with benefits outweighing costs by 3.5 to 20.5 times.

\bibliographystyle{ecta-fullname}
\bibliography{mybib2.bib}

\end{document}